%% file: DCEE_for_AOD_manuscript.tex
\pdfoutput=1

\documentclass[letterpaper,journal]{IEEEtran}

\usepackage{algorithm2e}
\usepackage{makecell}

\usepackage{array}
\usepackage{booktabs}
\usepackage{multirow}

\usepackage[caption=false,font=normalsize,labelfont=sf,textfont=sf]{subfig}
\usepackage{graphicx}
\usepackage{stfloats}
\usepackage{tikz}
\usepackage{pgfplots}
\pgfplotsset{compat=1.18}

\usetikzlibrary{3d,calc}
\usetikzlibrary{shapes.geometric, arrows}
\tikzstyle{block} = [rectangle, rounded corners, minimum height=3cm, text centered, text width=4cm, draw=black, fill=gray!10]
\tikzstyle{subblock} = [rectangle, rounded corners, minimum width=3.5cm, minimum height=0.9cm, text centered, draw=black, fill=white]
\tikzstyle{topic} = [ellipse, minimum width=3.2cm, minimum height=1.2cm, text centered, draw=black, fill=blue!10]
\tikzstyle{line} = [draw, thick, dashed, <->] %

\usepackage{amssymb}
\usepackage{amsmath,amsfonts}
\usepackage{xcolor}

\usepackage{cite}
\usepackage{lineno}

\usepackage{url}
\usepackage{verbatim}
\usepackage{textcomp}

\usepackage{hyperref}

\usepackage{robotarm}

\begin{document}

\title{
A Goal-Oriented Approach for Active Object Detection with Exploration-Exploitation Balance
}

\author{
Yalei~Yu,
Matthew~Coombes,
Wen-Hua~Chen, \IEEEmembership{Fellow, IEEE},
Cong~Sun,
Myles~Flanagan,
Jingjing~Jiang,
Promod~Pashupathy,
Masoud~Sotoodeh-Bahraini,
Peter~Kinnell,
and Niels~Lohse
\thanks{This work was supported by the UK Engineering and Physical Sciences Research Council (EPSRC) Established Career Fellowship “Goal-Oriented Control Systems: Disturbance, Uncertainty and Constraints” under the grant number EP/T005734/1, and EPSRC research grant “Industrial Robots-as-a-Service (IRaaS) - Resilient and responsive manufacturing systems enabled by rapidly deployable mobile robots" under the grant number EP/V050966/2. (Corresponding author: Wen-Hua Chen.)
    }
\thanks{Y. Yu, M. Coombes, W.-H. Chen, P. Pashupathy, and J. Jiang are with the Department of Aeronautical and Automotive Engineering, Loughborough University, Loughborough, LE11 3TU, U.K.
(email: \{y.yu2, m.j.coombes, w.chen, p.pashupathy, j.jiang2\}@lboro.ac.uk) }  
\thanks{C. Sun is with Adaptive \& Intelligent Robotics Lab, Imperial College London, London,  SW7 2AZ U.K.
    (email: c.sun1@imperial.ac.uk) }
\thanks{M. Flanagan and P. Kinnell are with Intelligent Automation Centre, Loughborough University, Loughborough, LE11 3TU U.K.
        (email: m.flanagan@gmail.com, p.kinnell@lboro.ac.uk) }
\thanks{M. Sotoodeh-Bahraini and N. Lohse are with Birmingham Institute for Robotics, University of Birmingham, College of Engineering and Physical Sciences, School of Engineering, Birmingham, B15 2TT U.K.
        (email: \{s.m.sotoodehbahraini, n.lohse\}@bham.ac.uk) }
}

\markboth{IEEE Transactions on Control Systems Technology
} 
{Yu \MakeLowercase{\textit{et al.}}: A Goal-Oriented Approach for Active Object Detection with Exploration-Exploitation Balance
}

\maketitle

\begin{abstract}
Active object detection, which aims to identify objects of interest through controlled camera movements, plays a pivotal role in real-world visual perception for autonomous robotic applications, such as manufacturing tasks (\emph{e.g.}, assembly operations) performed in unknown environments. 
This research focuses on optimally guiding a camera to identify objects of interest with a high confidence score, while simultaneously reducing the number of observations required to identify the best viewpoint.
A dual control for exploration and exploitation (DCEE) algorithm is presented within goal-oriented control systems to achieve efficient active object detection, leveraging active learning by incorporating variance-based uncertainty estimation in the cost function.
Specifically, active object detection is achieved through the development of a reward function that encodes knowledge about the confidence variation of objects as a function of viewpoint position within a given domain.
By identifying the unknown parameters of this function, the algorithm optimizes the camera viewpoint trajectory.
DCEE integrates parameter estimation of the reward function and view planning, ensuring a balanced trade-off between the exploitation of learned knowledge and active exploration during the planning process.
{Moreover, it adapts across the tested scenarios, from LEGO brick detection with a single obstacle to the detection of an industrial peg under two obstacles, without modification to the algorithm, its configuration settings, or its number of parameters.}
To validate the proposed approach, extensive numerical studies, high-fidelity virtual simulations, and real-world experiments were conducted.
The results confirm the effectiveness of DCEE, showcasing superior performance compared to existing methods, including model predictive control (MPC), entropy, {and upper confidence bound (UCB) acquisition} approaches.
\end{abstract}


\section{Introduction}
\IEEEPARstart{A}{ctive} object detection is a critical capability for autonomous robots tasked with executing operations in unknown environments with unspecified targets (or so-called references) \cite{jayaratne2019unsupervised}, such as object detection \cite{wu2024enhanced}, pose estimation \cite{zeng2020view}, and 3-D reconstruction \cite{cristofalo2020vision}.
In many instances, not all poses (\emph{i.e.}, viewpoints) of a vision sensor relative to the object of interest will allow the same level of identification certainty. 
This is especially the case in cluttered scenes or if objects have identifying characteristics that are not visible from all poses.
This task can be conceptualized as a form of action selection, as discussed in \cite{koerding2006bayesian}, viewpoint selection \cite{bonaventura2018survey}, viewpoint planning \cite{zeng2020view}, 
or next-best-view determination \cite{doumanoglou2016recovering}, distinguishes itself from passive object detection by enabling the exploration of unknown environments.
Specifically, active object detection involves dynamically adjusting the viewpoint of a camera mounted on a robot to gather relevant visual information for identifying objects of interest by finding a good pose.
It is often not obvious what this pose is in an unseen scenario.
Random or exhaustive searches can be very effort-intensive.
Hence, the aim is to minimize data acquisition while adhering to operational constraints \cite{ammirato2017dataset}.

Various methods for active object detection have been investigated, including information-theoretic approaches \cite{sock2017multi, bonaventura2018survey, arbel1999viewpoint, denzler2002information, huber2012bayesian} and learning-based techniques \cite{hinterstoisser2011multimodal, tang2012textured, fang2021enhancing, xu2021towards, fang2022self}. 
Information-theoretic methods, which primarily rely on Shannon entropy to compute viewpoint entropy (\emph{i.e.}, information), have been widely used to quantify information. 
These methods aim to generate optimal view planning for detecting objects of interest by maximizing information gain \cite{bonaventura2018survey, sock2017multi, arbel1999viewpoint}. 
To establish the relationship between system states and actions, Bayesian inference and entropy are often integrated \cite{denzler2002information, huber2012bayesian}.
However, entropy methods predominantly emphasize the exploration of the object of interest and its surrounding environment, rather than the exploitation of the knowledge that has been learned. 
Consequently, achieving a balance between exploration and exploitation while controlling the camera to identify the best viewpoint remains a significant challenge.

On the other hand, learning-based methods primarily focus on leveraging trained models to detect objects and generate the subsequent camera motion. 
These methods have been extensively explored in active object detection scenarios, including template-based approaches \cite{hinterstoisser2011multimodal}, matching-based techniques \cite{tang2012textured, kriegel2013combining}, and statistical-learning-based methods \cite{fang2021enhancing, xu2021towards, fang2022self}.
{To increase the learning capabilities, some active learning methods have been designed by introducing a distribution-shattering strategy in \cite{cao2022shattering}, and a multi-head mechanism in \cite{li2024al}. }
However, due to the inherent characteristics of learning-based algorithms, these methods are limited in their ability to explore objects and uncertain environments that were not included in the training data \cite{hinterstoisser2011multimodal}. 
In other words, learning-based approaches exhibit significant drawbacks, such as their reliance on large datasets for training and limited flexibility in adapting to different scenarios in real-time applications \cite{hinterstoisser2011multimodal, fang2022self}.
In this context, learning-based methods also suffer from poor flexibility, making it challenging to extend their application to objects or environments that were not part of the original training set.

To address the aforementioned challenges, a promising approach for naturally balancing exploration and exploitation is the dual control for exploration and exploitation (DCEE) within goal-oriented control systems (GOCS), as proposed in \cite{chen2024goal} and \cite{chen2021dual}. 
DCEE represents a strategy for achieving GOCS by automatically identifying optimal manipulation conditions, thereby enhancing the level of automation. 
Specifically, DCEE integrates exploration (\emph{i.e.}, probing the target and its environment) and exploitation (\emph{i.e.}, tracking the estimated object reference) within a unified framework. 
As a result, DCEE inherently incorporates active learning capabilities for exploring unknown environments, distinguishing it from traditional active learning methods \cite{fu2019active}.
{Similar active learning methods have been applied for control applications \cite{zhang2021trajectory} and object recognition \cite{nie2024online}.}  
Notably, DCEE, as outlined in \cite{chen2021dual}, does not rely on extensive datasets and performs effectively in time-varying environments. 
This characteristic underscores its advantages over machine learning-based methods \cite{fang2021enhancing}, particularly in terms of adaptability and efficiency.
Subsequently, DCEE has been successfully applied in various domains, including autonomous search \cite{li2025cooperative,tan2024multistep,rhodes2022autonomous}, maximal power point tracking \cite{yu2026k}, anti-lock emergency braking \cite{sullivan2023exploration}, and object tracking \cite{glover2023dual}.

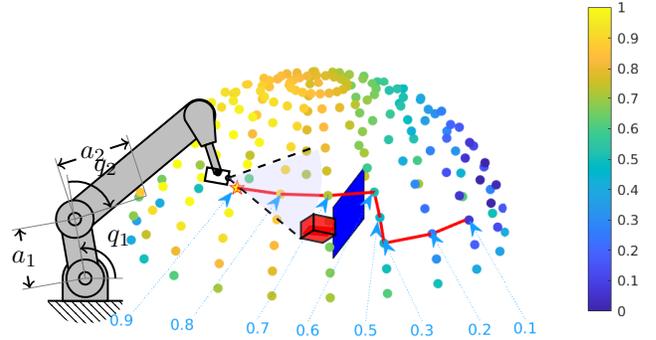
\begin{figure}
\centering
\input{Figures/Demo_AOD_DCEE.tex}
\caption{An example of active object detection using a robot arm equipped with a camera is illustrated. The red star marks the terminal position, where the confidence score of the object of interest (\emph{i.e.}, the red cube) exceeds a predefined threshold (\emph{e.g.}, 0.9). The blue area represents an obstacle obstructing the cube from one side. The numerical values at each position indicate the corresponding confidence scores of the object.
    }
    \label{fig_setting_example_aod}
\end{figure}

Active object detection within goal-oriented control systems can be formulated as an optimization problem, where the objective is to identify the optimal viewpoint within a given domain. 
If a function is available to quantify confidence scores across different viewpoints, the optimal viewpoint can be identified by optimizing this function.
The DCEE algorithm addresses this challenge by estimating the function and guiding the system toward the optimal viewpoint. 
This process is illustrated in Fig. \ref{fig_setting_example_aod}, where the color bar represents variations in confidence levels (\emph{i.e.}, the function's output) across the domain. 
As the function is progressively estimated, the camera transitions from low-confidence regions to high-confidence areas, following the optimized trajectory depicted in red.
By effectively balancing exploration (\emph{i.e.}, identifying objects in unknown environments) and exploitation (\emph{i.e.}, tracking the estimated optimal viewpoint), DCEE provides a more comprehensive solution compared to methods that prioritize only one aspect.
In contrast, entropy-based approaches, such as those in \cite{bonaventura2018survey}, focus solely on exploration, while learning-based techniques, as discussed in \cite{xu2021towards} and \cite{fang2022self}, primarily emphasize exploitation.

Motivated by the above observation, the contributions of this paper can be summarized as follows: 
\begin{itemize}
        \item The maximal cost function-based DCEE algorithm is developed and applied for the first time in the context of active object detection. 
        This algorithm inherently integrates both exploration, which involves discovering objects in unknown environments, and exploitation, which optimizes the camera's viewpoint based on an estimated confidence score. 
        The exploration component is specifically characterized by the introduction of variance-based uncertainty estimation in the cost function, which actively guides the search for the optimal viewpoint trajectory, thereby enhancing object detection performance in unfamiliar environments.
        \item The reward function, formulated as a linear regression model, is designed to encode prior information about object confidence as a function of viewpoint positions. 
        This formulation enables the reward function to effectively represent the spatial distribution of confidence scores using only six parameters. 
        The designed model is validated across {four} distinct scenarios, {including a different object under a more severely occluded, two-obstacle scene,} consistently maintaining the same number of parameters.
        {This consistency indicates that the model can capture confidence variations across the environments considered.}
        \item The expected performance of the DCEE algorithm is validated through numerical and high-fidelity virtual simulations, with its effectiveness further confirmed by real-world experiments, highlighting both its adaptability {across the tested environments} and its practical utility in engineering applications.
        \item To highlight the advantages of the DCEE algorithm,
        quantitative analyses are performed, comparing DCEE with MPC and entropy methods, where MPC is introduced for object detection for the first time in this paper, {and with a further UCB acquisition baseline reformulated within the present setting}.
        The results demonstrate the superior performance of DCEE, particularly in terms of average convergence distance and variance reduction in parameter estimation. {Moreover, unlike the UCB baseline, DCEE does not require a tunable coefficient to balance exploration and exploitation.}
\end{itemize}

\newtheorem{assumption}{Assumption}
\newtheorem{remark}{Remark}
\newtheorem{define}{Definition}
\newtheorem{theorem}{Theorem}

\begin{define}
  The following notations are used in this paper. 
  $ \mathbb{R}^m $ represents the $m$-dimensional Euclidean space.
  $ {\left( \cdot \right)^{\rm T}} $ denotes the transpose of a matrix $ {\left( \cdot \right)} $, while $\left\| \cdot \right\| $ represents the Euclidean norm. 
  $(\widetilde\cdot)$ describes the estimation error. 
\end{define}

\section{Problem formulation}
This section will sequentially provide the system modeling, reward function for environmental awareness, sensor modeling, and research objectives.

\subsection{System modeling}
The movement of a camera \cite{jayaraman2016look} is modeled as follows:
\begin{equation}\label{eq_agent_model_aod}
{\rm {\mathrm{p}}}_{k+1}= {\rm {\mathrm{p}}}_{k}+u_k
\end{equation}
where $ {{\mathrm{p}}}_{k}=[{{\mathrm{p}}}_{k,x} ~ {{\mathrm{p}}}_{k,y} ~ {\mathrm{p}}_{k,z}]^{\rm T} \in {\Omega} \subseteq \mathbb{R}^3$ is the position of the camera at the current time step $k$, with $ \Omega$ being the domain of operating environments. 
For convenience, ${{\mathrm{p}}}_{k}$ is simplified as ${{\mathrm{p}}}_{k}=[{{\mathrm{p}}}_{x} ~ {{\mathrm{p}}}_{y} ~ {\mathrm{p}}_{z}]^{\rm T}$.
The term $u_k \in \mathcal{U} \subseteq \mathbb{R}^3$ denotes the control action, with $\mathcal{U}$ being the admissible set of actions. 
Throughout, the sets $\Omega$ and $\mathcal{U}$ are taken to be closed. 

\subsection{Reward function for environment awareness}
The reward function, denoted as ${C}({\mathrm{p}}_{k}, \theta)$, establishes the relationship between the confidence score of objects and the camera's viewpoint position. 
It represents the confidence associated with the object of interest (also referred to as the target or reference) in unknown environments. 
This relationship can be expressed through a linear regression model as follows:
\begin{equation}\label{eq_aod_model_sensor}
    C({\mathrm{p}}_{k}, \theta) =  \phi({\mathrm{p}}_{k})^{\rm T}\theta  
\end{equation}
where $ {\mathrm{p}}_{k} \in \mathbb{R}^{3} $ denotes the outputs of the system given by (\ref{eq_agent_model_aod}). 
The terms $ \theta \in \mathbb{R}^{m} $ and $ \phi({\mathrm{p}}_{k}) \in  \mathbb{R}^m $ represent the vector of unknown true parameters associated with the object and its environment, and the basis function (\emph{i.e.}, the regressor) of the model, respectively. 

\subsection{Sensor modeling}
Suppose that at each time step, the system output ${\mathrm{p}}_{k} $ and the reward function $C({\mathrm{p}}_{k}, \theta)$ can be measured or derived, subject to measurement noise.
According to the reward function for environment sensing given by (\ref{eq_aod_model_sensor}), the point-wise measurement at the moment $k$ is defined as follows: 
\begin{equation}\label{eq_meas_model}
   \mathcal{C}_k =
   \left\{
   \begin{aligned}
     & C({\mathrm{p}}_{k}, \theta)+\mu_k, & D=1 \\
    & \mu_k, & D=0
   \end{aligned}
   \right.
\end{equation}
where $D$ denotes a detection event and $D=1$ denotes successful detection while $D=0$ represents missed detection.
The term $ \mu_{k} \in \mathbb{R} $ denotes measurement noise imposed on sensor readings and data processing.
{In implementation, a missed detection or an incorrect classification is recorded as a zero confidence score, so that the measurement reduces to noise alone ($D=0$).}

\begin{remark}\label{rem_measure_model}
The reward function for environment awareness, as defined in (\ref{eq_aod_model_sensor}), is developed to encode knowledge about the variation in confidence score as a function of viewpoint position within a given domain. 
In this context, the proposed function plays a role analogous to the entropy map in \cite{arbel1999viewpoint}, which characterizes the relationship between object discriminability and viewing position. 
It also resembles the information content map in \cite{takeuchi1998active} and the uncertainty map in \cite{wang2011bayesian}, both of which describe the spatial distribution of information content. 
Moreover, such reward functions have been widely adopted in various applications, for instance, in the spatial maps used for source seeking \cite{mellucci2020environmental,hutchinson2019information}, which represent spatial phenomena as a function of location.

Specifically, in the reward function (\ref{eq_aod_model_sensor}), $\theta$ represents the unknown parameter to be estimated, which is associated with the object of interest (\emph{i.e.}, the target) and the unknown environment in which it is located.
The basis function $\phi({\mathrm{p}}_{k})$ is a smooth function of ${\mathrm{p}}_{k}$. 
There are two primary approaches to determining the structure of $\phi({\mathrm{p}}_{k})$.
The first involves first-principle modeling, as demonstrated in source term estimation methods described in \cite{hutchinson2020unmanned}.
The second approach employs function approximation techniques such as those used in maximal power point tracking reported in  \cite{yu2026k}.
From another perspective, the form of the reward function given by (\ref{eq_aod_model_sensor}) is widely employed in inverse reinforcement learning, as discussed in \cite{arora2021survey} and \cite{zare2024survey}.
\end{remark}

\begin{remark}\label{rmk_meas_errors}
It is important to note that the measurement $\mathcal{C}_k
$ does not correspond to direct sensor readings, such as raw images captured by cameras. 
Instead, it represents the confidence score of the object of interest, which is computed through image processing using well-trained YOLOv5-s models in \cite{jocher2022ultralyticsyolov5} applied to images captured at the current time step $k$.
{In this context, it should be noted that these measurements are sporadic and intermittent due to factors such as object occlusion and environmental influences (\emph{e.g.}, lighting conditions).
}
\end{remark}

\begin{assumption}\label{assm_aod_mapping_opti_reward}
    There exists a smooth function $g(\cdot)$: $\mathbb{R}^{m} \rightarrow \mathbb{R}^3 $ such that 
\begin{equation}\label{ass_first_ord_diff}
      \left.  \frac{\partial C({\mathrm{p}_k}, \theta)}{\partial {\mathrm{p}_k}} \right|_{\mathrm{p}_k = \mathrm{p}^*} = 0 ~ \textit{if and only if} ~ {\mathrm{p}}^* = g(\theta ).
\end{equation}
\end{assumption}

\begin{assumption}\label{assm_aod_exist_opti_reward}
   The reward function $ C({\mathrm{p}}_{k}, \theta) $ is assumed to be twice differentiable and strictly {concave} on $ {\mathrm{p}}_{k} $ for any $ \theta \in \mathbb{R}^m $ as given below:
   \begin{align}
       \frac{\partial^2 C({\mathrm{p}}_k, \theta)}{\partial {\mathrm{p}_k} \partial {\mathrm{p}}_k} <  0.
   \end{align}
\end{assumption}

\begin{assumption}\label{assm_aod_noise}
The measurement noise $\mu_{k}$ is assumed to be independent and identically distributed with zero mean and bounded variance, expressed mathematically as: $\mathbb{E}[\mu_{k}]= 0$ and $\mathbb{E}[\mu_{k}^2] \leqslant \sigma^2$.
\end{assumption}

\begin{remark}
   Assumption \ref{assm_aod_mapping_opti_reward} posits that the reward function $ C(\mathrm{p}_k, \theta) $ attains optimality, if and only if the system reaches the optimal output equilibrium $ \mathrm{p}_k = \mathrm{p}^* = g(\theta) $ \cite{tan2009global}.
   Assumption \ref{assm_aod_exist_opti_reward} is established to ensure that the function $ C(\mathrm{p}_k, \theta) $ satisfies the {concavity} property.
   In Assumption \ref{assm_aod_noise}, the noise $\mu_k$ can follow any probability distribution and is not limited to Gaussian noise.
\end{remark}

\begin{remark}\label{rmk_scope_concavity}
{Assumption \ref{assm_aod_exist_opti_reward} constrains the reward model used for planning, not the physical confidence field, which is generally not concave. By the measurement model (\ref{eq_meas_model}), occlusion causes detection failure, reducing the measurement to noise and potentially introducing non-concavity at occlusion-shadow boundaries.
Assumptions \ref{assm_aod_mapping_opti_reward} and \ref{assm_aod_exist_opti_reward} therefore restrict attention to scenarios in which a single object of interest admits a unique optimal viewpoint, covering geometrically complex objects and multiple occlusions.
Moreover, strict concavity is not essential and can be relaxed to quasiconcavity, provided that the reward retains a unique optimum over the domain. An example is the isotropic plume model adopted for source term estimation in \cite{chen2021dual}, in which concentration decreases monotonically with distance from the source.
DCEE is therefore applicable to both concave and quasiconcave reward models.}
\end{remark}

\subsection{Research objectives}
\label{sec_research_obj}
Before the formulation of the research objective, we introduce some notations for convenience. 
The belief regarding the unknown parameters $\theta$ in the reward function for the confidence score of the object of interest, at time step $k$, can be characterized by the probability density function $p(\theta_k)$.
The information state is denoted as $ \mathcal{Z}_k=[\mathrm{p}_{k}; u_{k-1}; \mathcal{C}_{k}] $.
The accumulated measurements up to time step $k$ are denoted as $ \mathfrak{C}_k= \{ \mathcal{Z}_1, \mathcal{Z}_2,\dots,  \mathcal{Z}_k\}$. 
Then the posterior distribution of the parameters at time step $k$ is defined as $\rho_{k|k}:= p(\theta_k | \mathfrak{C}_k)$. 
Control action guiding the agent to a new position, ${ \mathrm{p}}_{k+1 \vert k}$, results in the upcoming confidence measurement, $ \hat{\mathcal{C}}_{k+1 \vert k} = \hat{\mathcal{C}}_{k+1|k}(\mathrm{p}_{k+1|k}, \theta_{k+1|k}) $, which can be treated as a random variable, denoted as $ \hat{\mathcal{C}}_{k +1 \vert k}\sim p(\hat{ \mathcal{C}}_{k+1 \vert k}\vert  u_k)$.
Based on the anticipated future measurement, a hypothetical posterior distribution of the parameter is derived as $\rho_{k+1|k}:= p(\theta | \mathfrak{C}_k, \mathcal{Z}_{k+1|k})$, where $\mathcal{Z}_{k+1|k} = [\mathrm{p}_{k+1|k}; u_k; \hat{\mathcal{C}}_{k+1|k}]$. 
Consequently, the control action influences both the future state and the belief in the parameter estimation. 
Then the research objective can be formulated as follows:
\begin{align}
\label{eq_cost_fun_research_obj}
& \max_{u_k \in \mathcal{U}} J({u_k}) =  \max_{u_k \in \mathcal{U}}  \mathbb{E}[ \hat{\mathcal{C}}_{k+1 \vert k}^2({\mathrm{p}}_{k+1|k},\theta_{k+1|k} ) | \mathfrak{C}_{k+1|k} ]
\end{align}
subject to (\ref{eq_agent_model_aod}), $ \mathrm{p}_{k+1|k} \in \Omega $ and $ u_k \in \mathcal{U} $.
The term $ \mathfrak{C}_{k+1|k} $ is defined as $\mathfrak{C}_{k+1|k}=\{\mathfrak{C}_{k}, \mathcal{Z}_{k+1|k} \}  $.

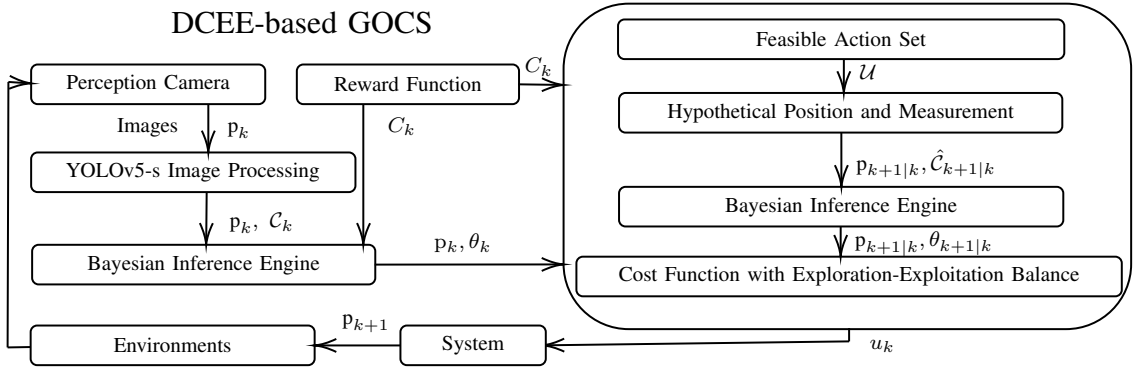
\begin{figure*}
\centering  
\input{Figures/AOD_flow_chart}  
\caption{Structure of DCEE-based goal-oriented control systems (GOCS) for active object detection, where $C_k = C(\mathrm{p}_k,\theta)$.
 } 
\label{dia_dcee_gocs}
\end{figure*}

\section{DCEE algorithm design}
This section outlines the formulation of the DCEE algorithm within goal-oriented control systems and its implementation using a Bayesian inference engine. 
The corresponding structure is illustrated in Fig. \ref{dia_dcee_gocs}.

\subsection{DCEE approach design} \label{subsec_aod_dcee_formu} 
The control variable $u_k$ is designed to direct the camera to a position where the predicted posterior measurement $\hat{\mathcal{C}}_{k+1 \vert k}$ is maximal. 
In this context, given the research objective (\ref{eq_cost_fun_research_obj}), the cost function $J(u_k) $ can be formulated as follows:
\begin{align}
\label{eq_cost_fun_1}
& \max_{u_k \in \mathcal{U}} J({u_k}) \nonumber \\
= & \max_{u_k \in \mathcal{U}}  \mathbb{E}[\hat{\mathcal{C}}_{k+1 \vert k}^2({\mathrm{p}}_{k+1|k},\theta_{k+1|k}) | \mathfrak{C}_{k+1\vert k} ] \nonumber   \\ 
= & \max_{u_k \in \mathcal{U}}  \left\{   \bar{\mathcal{C}}_{k+1 \vert k}^2({\mathrm{p}}_{k+1|k},\theta_{k+1|k}) \right. \nonumber \\ 
& \left. + \mathbb{E}[\widetilde{\mathcal{C}}_{k+1 \vert k}^2({\mathrm{p}}_{k+1|k},\theta_{k+1|k}) ] \right\} 
\end{align}
subject to (\ref{eq_agent_model_aod}), $ \mathrm{p}_{k+1|k} \in \Omega $ and $ u_k \in \mathcal{U} $.
The terms in (\ref{eq_cost_fun_1}) are given as $\bar{\mathcal{C}}_{k+1 \vert k}({\mathrm{p}}_{k+1|k},\theta_{k+1|k})= \mathbb{E}[\hat{\mathcal{C}}_{k+1 \vert k}({\mathrm{p}}_{k+1|k},\theta_{k+1|k})]$ and $\widetilde{\mathcal{C}}_{k+1 \vert k}({\mathrm{p}}_{k+1|k},\theta_{k+1|k})=\hat{\mathcal{C}}_{k+1 \vert k}({\mathrm{p}}_{k+1|k},\theta_{k+1|k})-\bar{\mathcal{C}}_{k+1 \vert k}({\mathrm{p}}_{k+1|k},\theta_{k+1|k})$, and thus $ \mathbb{E}[ \widetilde{\mathcal{C}}_{k+1 \vert k}({\mathrm{p}}_{k+1|k},\theta_{k+1|k})] =0 $. Then the cost function given by (\ref{eq_cost_fun_1}) for the optimization problem can be simplified as follows: 
\begin{align}\label{eq_costfunction_juk}
 \max_{u_k \in \mathcal{U} } J({u_k}) 
= &  \max_{u_k \in \mathcal{U} }  [\bar{\mathcal{C}}_{k+1 \vert k}^2({\mathrm{p}}_{k+1|k},\theta_{k+1|k}) \nonumber \\
& +\mathcal{P}_{k+1 \vert k}({\mathrm{p}}_{k+1|k},\theta_{k+1|k}) ]
\end{align}
subject to (\ref{eq_agent_model_aod}), $ \mathrm{p}_{k+1|k} \in \Omega $ and $ u_k \in \mathcal{U} $.
The term $ \mathcal{P}_{k+1 \vert k}({\mathrm{p}}_{k+1|k},\theta_{k+1|k}) = \mathbb{E}[\widetilde{\mathcal{C}}_{k+1 \vert k}^2({\mathrm{p}}_{k+1|k},\theta_{k+1|k}) ] = {\rm cov}\{ \widetilde{\mathcal{C}}_{k+1|k} ({\mathrm{p}}_{k+1|k}, \theta_{k+1|k}) \} = {\rm cov}\{ \hat{\mathcal{C}}_{k+1|k} ({\mathrm{p}}_{k+1|k}, \theta_{k+1|k}) \} $. 

\begin{remark}
    The cost function of DCEE, as defined in (\ref{eq_costfunction_juk}), consists of two key components. The first, the exploitation term $ \bar{\mathcal{C}}_{k+1 \vert k}^2({\mathrm{p}}_{k+1|k},\theta_{k+1|k}) $ is responsible for utilizing learned knowledge to optimize the camera's viewpoint. The second, the exploration term $ \mathcal{P}_{k+1 \vert k}({\mathrm{p}}_{k+1|k},\theta_{k+1|k}) $ facilitates the discovery of new information by guiding the exploration of objects in previously unknown environments.
    In this case, DCEE naturally integrates exploration and exploitation functions, thus achieving an exploration-exploitation balance.
\end{remark}

\begin{remark}\label{rmk_max_dcee}
{Note that the designed cost function in (\ref{eq_costfunction_juk}) differs from the existing DCEE formulations presented in \cite{chen2021dual} and \cite{li2025cooperative}. 
In the existing DCEE algorithms \cite{chen2021dual, li2025cooperative}, the objective is to minimize the cost function by reducing tracking errors between the current position and the estimated reference, while simultaneously minimizing estimation errors of the unknown parameters. 
The cost function of DCEE, as defined in (\ref{eq_costfunction_juk}), aims to maximize the cost function. 
Specifically, it seeks to maximize the believed confidence score and the believed uncertainties associated with exploring unknown areas.
In this way, the learned knowledge is fully utilized, while moving into the most uncertain areas further enhances exploration and learning outcomes.
These results are subsequently incorporated into the exploitation term, thereby reducing uncertainties through active learning.
Note that this is the key distinction between DCEE, which accounts for variance, and existing optimization methods \cite{rawlings2020model}.
Moreover, this method of maximizing the cost function has been applied to autonomous search tasks reported in \cite{tran2024fast}, achieving the expected performance.
Similarly, a variance-based cost function with manually adjusted weights is proposed in \cite{sorg2012variance}.
In \cite{benciolini2025active}, exploration is directed toward the most uncertain regions by maximizing the posterior variance of a Gaussian process model.
{Unlike the method developed in this paper, which naturally balances exploration and exploitation without such tuning, \cite{li2018adaptive} requires the introduction of a regulator factor.}
Furthermore, the concept of maximizing uncertainty aligns with the principle of ``optimizing in the face of uncertainty'' in reinforcement learning, as discussed in \cite{bubeck2012regret}.}
\end{remark}

\subsection{Implementation of Bayesian inference engine}
The unknown parameter $ \theta$ in the reward function given by (\ref{eq_aod_model_sensor}) needs to be estimated by estimators.
In this context, the Bayesian estimator described in \cite{hutchinson2017review} is introduced to estimate the unknown parameters $ \theta $. 

The posterior distribution $ p( \theta_k | \mathfrak{C}_k ) $ mentioned in Subsection \ref{sec_research_obj} is subsequently updated through a Bayesian inference engine using the sensory data as follows:
\begin{align}
p( \theta_k | \mathfrak{C}_k ) =  \frac{p(\theta_k | \mathfrak{C}_{k-1})p(\mathcal{C}_{k} | \theta_k) }{ p( \mathcal{C}_k | \mathfrak{C}_{k-1}) }
\end{align}  
where 
\begin{equation}
p(\mathcal{C}_k | \mathfrak{C}_{k-1}) = \int	p(\mathcal{C}_k \vert \theta_k) p(\theta_k | \mathfrak{C}_{k-1}) {\rm d}\theta_{k}.
\end{equation}
The term $ p(\theta_k | \mathfrak{C}_{k-1}) $ denotes prior information of the unknown parameter $ \theta $ at the time step $k$.
If the prior information about the object of interest in the domain is available, it can be employed through an appropriate distribution to represent this prior knowledge. 
Otherwise, in the absence of prior information, the initial distribution $ p(\theta_0) $ can be specified as an uninformative prior such as a uniform distribution over the parameter domain bounds. 
 
The term $ p(\mathcal{C}_k|\theta_k) $ denotes the likelihood function, which is used to approximate the probability of the measured data $ \mathcal{C}_{k} $, given a hypothesized parameter estimate $ \theta_k $.  
In practical applications, the particle filter used in \cite{hutchinson2019information,hutchinson2020unmanned} is employed to implement a Bayesian inference engine, specifically for approximating the likelihood function.
Thus, the Bayesian estimator for the unknown parameters is constructed using the particle filter within the sequential Monte Carlo framework, as outlined in \cite{hutchinson2020unmanned}. 
The posterior distribution of the parameters, denoted as $p( \theta_k | \mathfrak{C}_k ) $, can then be approximated by a set of $N$ weighted samples $ \{ \theta_k^{i}, \omega_k^{i} \}^N_{i=1} $, expressed as follows:
 \begin{align}
     p( \theta_k | \mathfrak{C}_k )  \approx \sum^N_{i=1} \omega_k^i\delta ( \theta_k - \theta^i_k )
 \end{align}
where $ \delta(\cdot) $ represents a Dirac delta function, and $\theta_k^i $ is a sample representing a potential estimated parameter, and $\omega^i_k $ is the corresponding normalized weighting, satisfying $ \sum^N_{i=1}\omega_k^i =1 $.
For a detailed implementation of the particle filter, please refer to \cite{hutchinson2019information, hutchinson2020unmanned}.
The DCEE algorithm with Bayesian inference engine is given in Algorithm \ref{pseu_alg_dcee}.

\begin{remark}
The Bayesian inference engine is particularly well-suited for estimating uncertain parameters, as it excels in probabilistic frameworks when handling uncertain information, providing parameter estimates along with confidence levels.
The issue of sporadic and intermittent measurements of object confidence is discussed in Remark \ref{rmk_meas_errors}.
Furthermore, regarding its effectiveness, the Bayesian inference engine has been proven to be an unbiased estimation method, as demonstrated in \cite{li2022bayesian}. 
\end{remark}

\RestyleAlgo{ruled}
\SetKwComment{Comment}{/* }{ */}
\SetKwInOut{Initialize}{Initialize}
\SetKwInOut{Input}{Input}
\SetKwInOut{Output}{Output}
\SetKwComment{Set}{\textbf{\textrm{Set}}: }{}
\SetKwComment{Input}{\textbf{\textrm{Input}}: }{}
\SetKwComment{Output}{\textbf{\textrm{Output}}: }{}
\SetKwComment{Return}{\textbf{\textrm{Return}}: }{}

\begin{algorithm}[hbt!]
\caption{The DCEE algorithm}
\label{pseu_alg_dcee}
\Input{ \textrm{The basis function ${\phi}({\mathrm{p}_k})$ of reward function $C(\mathrm{p}_k,\theta) $, and the cost function $J(u_k)$ } }
\Initialize{The unknown parameters $\theta_{0} $, the system's initial state ${\mathrm{p}}_0$ }
\Set{ \textrm{ Feasible set of actions $ \mathcal{U}  $ and operation domain $\Omega$ } }
Estimate unknown parameters: 
$ p( \theta_k | \mathfrak{C}_k ) =  \frac{ p(\mathcal{C}_{k} | \theta_k) p(\theta_k | \mathfrak{C}_{k-1}) }{ p( \mathcal{C}_k | \mathfrak{C}_{k-1}) } $  \\
where $ p(\mathcal{C}_k | \mathfrak{C}_{k-1}) = \int	p(\mathcal{C}_k \vert \theta_k) p(\theta_k | \mathfrak{C}_{k-1}) {\rm d}\theta_{k} $ \\
Choose the next action $ u_k \in \mathcal{U}  $ by comparing the cost function: 
$  \max_{u_k \in \mathcal{U} } J({u_k}) 
= \max_{u_k \in \mathcal{U} }  [\bar{\mathcal{C}}_{k+1 \vert k}^2({\mathrm{p}}_{k+1|k},\theta_{k+1|k}) $   \\
 $ +\mathcal{P}_{k+1 \vert k}({\mathrm{p}}_{k+1|k},\theta_{k+1|k}) ] $ \\ 
 Update the next movement: 
 $ {\rm {\mathrm{p}}}_{k+1}= {\rm {\mathrm{p}}}_{k}+u_k $ \\
\Return{\textrm{$ \mathrm{p}_{k+1} $}}
\end{algorithm}

\begin{table*}
\centering
\caption{Primary feature comparison among different active object detection algorithms{, with related discussion in Remark~\ref{rmk_ucb_ts_compar}}}
\label{tab_compar_metrics}
\begin{tabular*}{0.85\textwidth}{@{\extracolsep{\fill}}cccccc}
\hline
Algorithms & Views & \thead{Exploitation \& \\ Exploration} & {\thead{Exploration \\ mechanism}} & Adaptability &  \thead{Computation \\ load}   \\
\hline
DCEE (\ref{eq_costfunction_juk}) & Single & Both & {Predicted variance (active)} &  Easy  & Medium  \\
MPC (\ref{eq_mpc})  & Single  & Exploitation & {--} & Easy   &  Medium  \\
Entropy (\ref{eq_entropy}) & Single & Exploration & {Information gain} & Easy  &  Medium  \\
Entropy \cite{sock2017multi} & Multiple & Exploration & {Information gain} & Easy  &  Medium  \\
{UCB (\ref{eq_ucb})} & {Single} & {Both} & {Current uncertainty (tuned $\beta$)} & {Easy} & {Medium}  \\
{Thompson sampling \cite{bakshi2023guts}} & {Single} & {Both} & {Posterior sampling} & {Easy} & {Medium}  \\
Learning-based methods \cite{xu2021towards} & Multiple  & {Exploitation} & {--} & Difficult   & High   \\
Learning-based methods \cite{ fang2022self} & Single  & {Exploitation} & {--} & Difficult   & High   \\
\hline
\end{tabular*}
\end{table*}

\subsection{Relationship with existing methods}
\label{sec_mpc_entropy}
To facilitate a comparative analysis with existing methods, including entropy and MPC, this section presents the specific formulations for each technique. 
Entropy-based methods are widely employed in active object detection, offering various approaches to measuring information gain \cite{denzler2002information, huber2012bayesian, sock2017multi}. 
A comprehensive survey of different entropy variations is provided in \cite{bonaventura2018survey}.
Building on the formulation of DCEE in Subsection \ref{subsec_aod_dcee_formu}, the viewpoint entropy (\emph{i.e.}, information gain) for active object detection can be reformulated as follows:
\begin{align}\label{eq_entropy}
\max_{u_k \in \mathcal{U}}J({u_k}) = &
\max_{u_k \in \mathcal{U}} [ H(\hat{\mathcal{C}}_{k+1|k}(\mathrm{p}_{k+1|k}, \theta_{k+1|k})|\mathfrak{C}_{k+1|k}) ] 
\end{align}
and 
\begin{align}
& H(\hat{\mathcal{C}}_{k+1|k}(\mathrm{p}_{k+1|k}, \theta_{k+1|k})| \mathfrak{C}_{k+1|k})  \nonumber \\
= &   - \sum_{}p(\hat{\mathcal{C}}_{k+1|k}(\mathrm{p}_{k+1|k}, \theta_{k+1|k}) | \mathfrak{C}_{k+1|k} ) \nonumber  \\
 & \times \log p(\hat{\mathcal{C}}_{k+1|k}(\mathrm{p}_{k+1|k}, \theta_{k+1|k}) | \mathfrak{C}_{k+1|k} )
\end{align}
subject to (\ref{eq_agent_model_aod}), $ \mathrm{p}_{k+1|k} \in \Omega $ and $ u_k \in \mathcal{U} $.
Thus, maximizing the entropy, as defined in (\ref{eq_entropy}), is employed to accomplish the object detection task, with the primary aim of the exploration of the object.
Note that the formulation of viewpoint entropy presented in (\ref{eq_entropy}) is closely related to that in \cite{vazquez2003automatic}, with modifications tailored to the specific setting of this paper. 
In this work, the predicted measurement based on collected information is utilized to quantify the information at each viewpoint, as opposed to the area-based approach adopted in \cite{vazquez2003automatic}.

On the other hand, based on the formulation of DCEE provided in Subsection \ref{subsec_aod_dcee_formu}, the object detection problem can also be addressed using MPC in \cite{rawlings2020model}, which can be expressed as follows: 
\begin{align}\label{eq_mpc}
\max_{u_k \in \mathcal{U}} J({u_k}) 
= & \max_{u_k \in \mathcal{U}}  \mathbb{E}[ \hat{\mathcal{C}}_{k+1 \vert k}^2({\mathrm{p}}_{k+1|k},\theta_{k}) | \mathfrak{C}_{k} ] \nonumber \\
= & \max_{u_k \in \mathcal{U}}  \bar{\mathcal{C}}_{k+1 \vert k}^2({\mathrm{p} }_{k+1|k},\theta_{k}) \nonumber \\
& + \mathcal{P}_{k|k}(\mathrm{p}_{k+1|k}, \theta_k)
\end{align}
subject to (\ref{eq_agent_model_aod}), $ \mathrm{p}_{k+1 \vert k} \in \Omega $ and $ u_k \in \mathcal{U} $.
The term $\bar{\mathcal{C}}_{k+1 \vert k}^2 $ contributes to exploitation.
In contrast, the uncertainty term $\mathcal{P}_{k|k}(\mathrm{p}_{k+1|k}, \theta_k)$ {is evaluated with the current parameter estimate $\theta_{k}$, which does not depend on $u_k$; it therefore does not anticipate the effect of the control action on future parameter estimation, so MPC lacks the active-learning (dual) effect of DCEE and acts primarily as exploitation.}
Notably, the application of MPC to object detection is introduced for the first time in this work, which represents a key contribution of this paper.

{As an additional baseline for comparison, viewpoint selection can also be posed as a Bayesian optimization problem with a UCB acquisition function \cite{meera2019obstacle}.
For a fair comparison, this baseline is reformulated within the setting of this paper: the Gaussian process surrogate is replaced by the reward model, with its parameters inferred by the same Bayesian inference engine as DCEE, so that the predicted confidence and its current uncertainty entering the acquisition are computed from the current posterior $\theta_k$, as in the MPC formulation (\ref{eq_mpc}).
The UCB baseline is accordingly given by
\begin{align}\label{eq_ucb}
\max_{u_k \in \mathcal{U}} J(u_k) = & \max_{u_k \in \mathcal{U}} \big[ \bar{\mathcal{C}}_{k+1|k}(\mathrm{p}_{k+1|k},\theta_{k}) \nonumber \\
& + \beta \sqrt{\mathcal{P}_{k|k}(\mathrm{p}_{k+1|k},\theta_{k})} \big]
\end{align}
subject to (\ref{eq_agent_model_aod}), $\mathrm{p}_{k+1|k} \in \Omega$ and $u_k \in \mathcal{U}$, which retains the canonical form of the acquisition, the predicted mean weighted against the current standard deviation by the tunable coefficient $\beta$.}

\begin{remark}\label{rmk_ucb_ts_compar}
Comparing DCEE, MPC, and entropy methods, it is evident that entropy prioritizes exploration, whereas MPC emphasizes exploitation.
In this context, both entropy and MPC can be regarded as special cases of DCEE.
This distinction also explains why DCEE outperforms entropy and MPC, as demonstrated by numerical simulation, high-fidelity simulations, and real-world experiments.
{The UCB baseline introduced in (\ref{eq_ucb}) also addresses the exploration-exploitation trade-off by combining the predicted confidence with its uncertainty through a tunable coefficient $\beta$. 
However, its uncertainty is evaluated based on the current estimate $\theta_k$, whereas DCEE evaluates the predicted uncertainty associated with $\theta_{k+1|k}$, which accounts for the anticipated effect of the candidate action on parameter estimation. %
In addition, the exploration-exploitation balance in DCEE follows directly from the expected squared-confidence objective and does not require an additional weighting coefficient.
Thompson sampling \cite{bakshi2023guts} provides an alternative exploration-exploitation strategy by sampling a realization of parameters from the posterior and selecting the action that maximizes the predicted confidence under the sampled model.
Unlike DCEE, however, uncertainty is incorporated implicitly through posterior sampling rather than explicitly within the action-selection objective. %
}
On the other hand, learning-based methods, {including reinforcement-learning-based active vision, can learn viewpoint-selection policies from training data or interactions with training environments}.
{A qualitative comparison analysis of these methods is provided in Table \ref{tab_compar_metrics}, evaluating key aspects such as the number of views at every time step, exploitation \& exploration features, adaptability to different scenarios, and computation load.
Furthermore, learning-based methods typically require offline training and may involve models with a large number of parameters, whereas DCEE identifies a low-dimensional, six-parameter reward model online from measurements collected during operation, as detailed later.}
\end{remark}

\begin{remark}
It is worth noting that the proposed DCEE framework for active object detection fundamentally differs from conventional visual servoing techniques, which mainly include image-based \cite{zhang2025image} and position-based visual servoing \cite{guo2024robust,guo2026raid}.
Compared with image-based visual servoing \cite{zhang2025image}, which controls camera motion by computing interaction matrices derived from image features and their corresponding depths, the DCEE framework eliminates the need for such complex computations, thereby simplifying the system design and implementation process.
In contrast to position-based visual servoing \cite{guo2024robust}, which relies on estimating the object’s pose to guide camera movements, DCEE dispenses with pose estimation, leading to improved computational efficiency.
Furthermore, unlike visual servoing approaches that utilize an object’s shape parameter space (\emph{e.g.}, shape contours) for control, as in \cite{jean2012robust}, DCEE optimizes the camera viewpoint based on the confidence space (\emph{i.e.}, confidence gradients) of the target object, obtained from the bounding boxes generated by the You-Only-Look-Once (YOLO) detector.
\end{remark}

\section{Simulation study for LEGO brick detection}
\label{sec_lego}
This section presents numerical simulations and high-fidelity online virtual simulations conducted in Isaac Sim with ROS2 for LEGO brick detection.
The simulation results are available in the video access link: \url{https://youtu.be/FdFXst8uWxc?si=aAL23uZYahkuoJzX}.

\subsection{Identification of the reward function}
\label{subsec_reward_funs}
This subsection identifies the reward function given by (\ref{eq_aod_model_sensor}) that characterizes the relationship between the confidence score of the object of interest, as determined through image processing, and the camera's viewing position.
For the linear regression-based reward function $ {C} ({\mathrm{p}_k},\theta) $ given by (\ref{eq_aod_model_sensor}), considering a third-order regression basis, it can be expressed as follows:
\begin{align}\label{eq_reward_func_cart_coord_17}
   {C}({\mathrm{p}_{k}}, \theta) = \phi({\mathrm{p}}_{k})^{\mathrm{T}}\theta
\end{align} 
where $ \phi({\mathrm{p}}_{k}) = [1, {\mathrm{p}}_x^3, {\mathrm{p}}_y^3, {\mathrm{p}}_z^3, {\mathrm{p}}_x{\mathrm{p}}_y{\mathrm{p}}_z, {\mathrm{p}}_x^2{\mathrm{p}}_y, {\mathrm{p}}_x^2{\mathrm{p}}_z, {\mathrm{p}}_y^2{\mathrm{p}}_z,  {\mathrm{p}}_y^2{\mathrm{p}}_x, \\ {\mathrm{p}}_z^2{\mathrm{p}}_x, {\mathrm{p}}_z^2{\mathrm{p}}_y, {\mathrm{p}}_x^2, {\mathrm{p}}_y^2, {\mathrm{p}}_z^2, {\mathrm{p}}_x{\mathrm{p}}_y, {\mathrm{p}}_y{\mathrm{p}}_z, {\mathrm{p}}_x{\mathrm{p}}_z, \mathrm{p}_x, \mathrm{p}_y, \mathrm{p}_z ]^{\rm T} $ and $ \theta = [\theta_1, \theta_2, \dots,  \theta_{19}, \theta_{20}]^{\text{T}} $.
Note that, according to equation (\ref{eq_reward_func_cart_coord_17}), a total of twenty parameters need to be identified. 
To achieve this, datasets will be collected to facilitate parameter identification and subsequently reduce the number of parameters. 

\begin{figure}
\centering
\input{Figures/Demo_s1_AOD_DCEE}
\caption{A virtual environment is set up in Isaac Sim, featuring a red  $ 2\times2$ LEGO brick, with an obstacle positioned on the negative side of the $y$-axis indicated by a green arrow pointing away from the obstacle (\emph{i.e.}, the white area). This setup is designated as Scenario 1 (S1).
}
\label{fig_dataset_env_lego_s1_aod}
\end{figure}
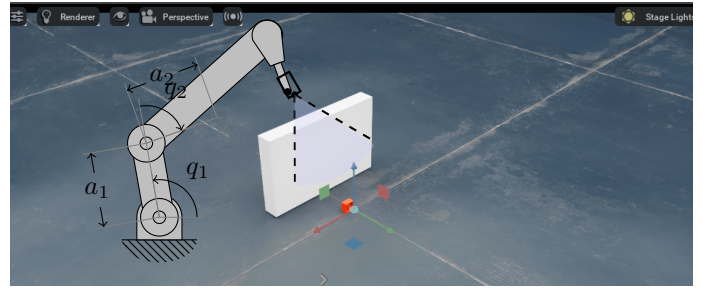

\begin{figure}
    \centering
    \includegraphics[trim=32 5 22 10, clip,width=0.5 \textwidth]{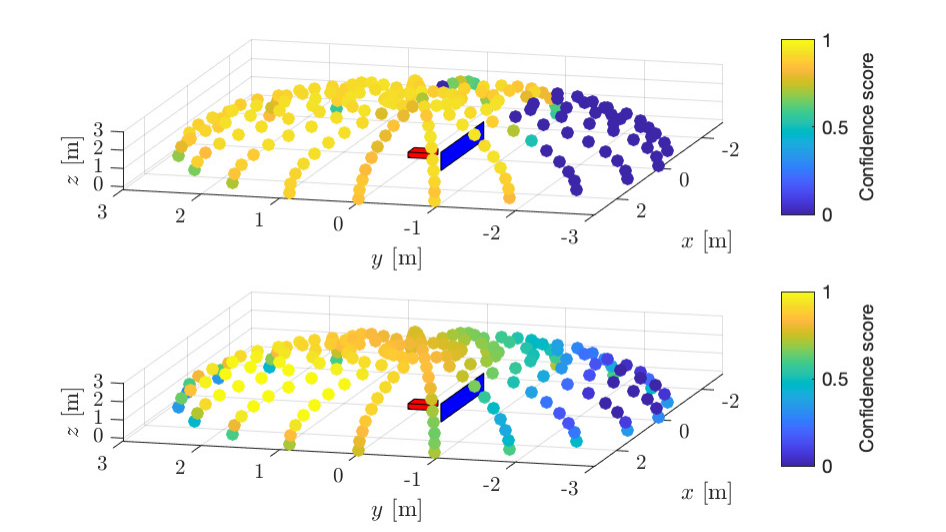}
    \caption{Collected datasets and the generated reward function with parameters listed in Table \ref{table_6para_carte_sys_lego_s1} illustrate the confidence scores of the red LEGO brick (\emph{i.e.}, the red cube) detected using the YOLOv5-s model for S1, as shown in Fig. \ref{fig_dataset_env_lego_s1_aod}. The data is acquired within a hemispherical domain, divided into an $11\times21$ grid.
    The blue area represents the obstacle.
    }
    \label{fig_datasets_mapping_lego_s1}
\end{figure}

\begin{figure}
    \centering
    \includegraphics[trim=32 5 22 10, clip, width=0.5 \textwidth]{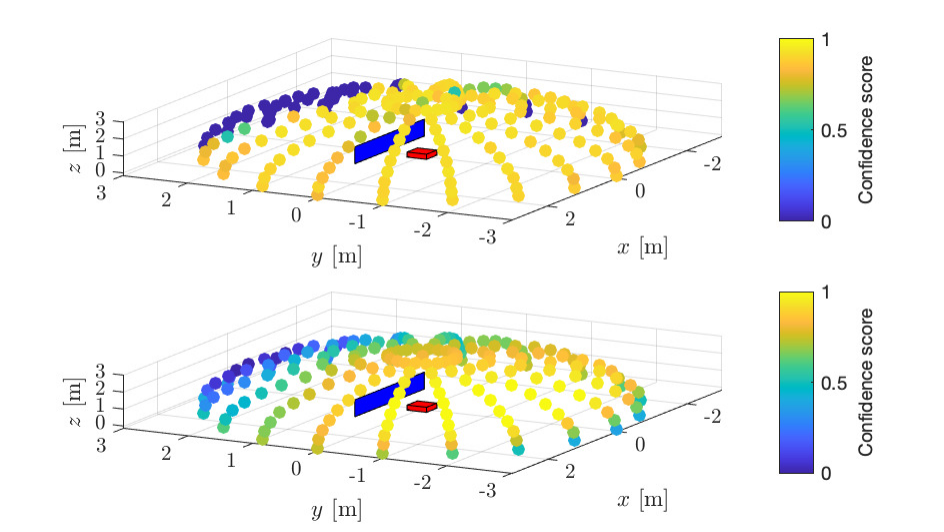}
    \caption{Collected datasets and the generated reward function with parameters listed in Table \ref{table_6para_carte_sys_lego_s2} illustrate the confidence scores of a red $ 2 \times 2$ LEGO brick for S2. 
    The data is acquired within a hemispherical domain, divided into an $11\times21$ grid.
    The blue area represents the obstacle.
    }
    \label{fig_datasets_mapping_lego_s2}
\end{figure}

\subsubsection{Collect datasets}
Considering the high performance of Isaac Sim in emulating various sensors, such as cameras \cite{audonnet2022systematic}, this high-fidelity virtual simulator is employed to represent the virtual environment. 
This virtual environment, as simulated in Isaac Sim, features the red $2\times2 $ LEGO brick (\emph{i.e.}, the object of interest) located at the origin of the domain, with an obstacle positioned on the negative side of the $y$-axis (indicated by a green arrow pointing away from the obstacle). 
This setup is illustrated in Fig. \ref{fig_dataset_env_lego_s1_aod} and is designated as Scenario 1 (S1).

A camera is then introduced into the virtual environment to capture images from various positions within a hemispherical domain, divided into an $11\times21$ grid. 
Throughout the process, the camera is assumed to face the origin of the domain consistently.
This assumption is also implemented in the online high-fidelity simulation conducted in Isaac Sim.
{This viewing sphere abstraction, in which candidate viewpoints are distributed over a sphere centered on the region of interest, is well established in viewpoint selection and next-best-view planning \cite{arbel1999viewpoint, sock2017multi, shi2025viso, lee2022uncertainty}.
Recent studies have adopted a similar hemisphere for active view planning in cluttered and occluded grasping scenes \cite{shi2025viso} and active 3D reconstruction \cite{lee2022uncertainty}, demonstrating the continued relevance of this abstraction across different applications.
Coupling camera orientation to position in this way serves three purposes:
it keeps the object of interest within the field of view throughout the domain, ensuring that a measurement is available whenever the object is not occluded; it makes the confidence depend on camera position alone, enabling the low-order reward model introduced in the following subsection;
and it reduces the viewpoint space to two dimensions, making the one-step optimization computationally tractable for real-time operation.}
These images are processed using YOLOv5-s model in \cite{jocher2022ultralyticsyolov5} fine-tuned with a custom LEGO assembly dataset. 
The dataset consists of $5184$ images with annotations generated in Isaac Sim. 
Once the confidence scores of the LEGO brick at different locations are obtained, the results are presented in the first sub-figure in Fig. \ref{fig_datasets_mapping_lego_s1}. 
In the following subsection, the collected datasets will be utilized to identify the reward model given by (\ref{eq_reward_func_cart_coord_17}).

\subsubsection{Parameter identification}
Based on the collected datasets for S1, as shown in Fig. \ref{fig_datasets_mapping_lego_s1}, using the least-squares method, the identified parameter values and the mean error are presented in Table \ref{table_para_carte_sys_17_1}. 
The mean error quantifies the average discrepancy between the predicted outputs of the identified model and the actual confidence scores obtained from the datasets.

Given the large number of parameters, some of which have relatively small values, a possible approach to simplifying the model given by (\ref{eq_reward_func_cart_coord_17}) is to exclude these smaller values, thereby reducing the number of parameters.
By implementing this approach, the number of parameters can be reduced from $20$ to $6$.
{
To examine the trade-off between model complexity and fitting accuracy, Table \ref{table_ablation_num_param} reports the mean and maximum fit error on the S1 dataset as the number of retained parameters is reduced from $20$ to $12$, $6$, and $4$.
The mean error increases from $0.1288$ to $0.1611$ when the model is reduced from $20$ to $6$ parameters, while further reduction to $4$ parameters increases the error to $0.1996$; the maximum error follows a similar pattern, remaining close to $0.75$ down to six parameters and rising to $0.81$ at four.
Six parameters are therefore selected as the compromise between model complexity and accuracy. 
Its applicability is further evaluated on the independent S2 dataset in Table \ref{table_6para_carte_sys_lego_s2} and subsequently through the high-fidelity and real-world studies.}
The remaining parameters are defined as new parameters $\theta_{1}, \theta_{2}, \theta_{3}, \theta_{4}, \theta_{5} $ and $ \theta_{6} $, associated with the corresponding regression basis $ {\mathrm{p}}_z^2 {\mathrm{p}}_y, {\mathrm{p}}_x^2, {\mathrm{p}}_y^2, {\mathrm{p}}_z^2, {\mathrm{p}}_y{\mathrm{p}}_z   $ and $ {\mathrm{p}}_x{\mathrm{p}}_z $, respectively.
As a result, the simplified third-order model, expressed in component form, is presented as follows:
\begin{align}\label{eq_6para_carte_coord_sys}
    {C}({\mathrm{p}_k}, \theta) = 
    \begin{bmatrix}
        {\mathrm{p}}_z^2 {\mathrm{p}}_y & {\mathrm{p}}_x^2 & {\mathrm{p}}_y^2 & {\mathrm{p}}_z^2 & {\mathrm{p}}_y{\mathrm{p}}_z & {\mathrm{p}}_x{\mathrm{p}}_z 
    \end{bmatrix}
    \begin{bmatrix}
        \theta_1 \\
        \vdots \\
        \theta_{6} 
    \end{bmatrix}
\end{align} 

The simplified model presented in (\ref{eq_6para_carte_coord_sys}) is then employed to identify the new unknown parameters, following the same methodology and datasets as before.
The identified parameter values are provided in Table \ref{table_6para_carte_sys_lego_s1}.
{Subsequently, based on the model defined by (\ref{eq_6para_carte_coord_sys}), the optimal viewpoint $\mathrm{p}^* = g(\theta)$ is obtained numerically by maximizing (\ref{eq_6para_carte_coord_sys}) over the operating domain $\Omega$, as implemented in the simulations and experiments.}

To evaluate the performance of the identified model, the obstacle in S1 is repositioned to the positive side of the $y$-axis, defining Scenario 2 (S2).
Following a similar approach to identifying the reward model given by (\ref{eq_6para_carte_coord_sys}) for S1, the collected dataset and the identified model for S2 are presented in Fig. \ref{fig_datasets_mapping_lego_s2}. 
The identified parameters for S2 are provided in Table \ref{table_6para_carte_sys_lego_s2}.

\begin{remark}
The identified parameters for S2 given in Table \ref{table_6para_carte_sys_lego_s2} demonstrate that the reward function defined by (\ref{eq_6para_carte_coord_sys}) performs effectively across different scenarios where the obstacle is repositioned to any location within the domain.
In other words, the identified model structure given by (\ref{eq_6para_carte_coord_sys}) remains valid regardless of the obstacle’s orientation relative to the LEGO brick.
Furthermore, the effectiveness of the model is later validated in real-world scenarios involving changes in both the brick's position and the obstacle's pose.
{Its adaptability and flexibility are further verified in Section~\ref{sec_peg_s4} for a different object, an industrial peg, under two obstacles (S4).
This performance indicates broader applicability, provided that the confidence score changes monotonically over the domain, decreasing away from a unique optimal viewpoint, as discussed in Remark~\ref{rmk_scope_concavity}.}
\end{remark}

\begin{table}
\caption{ Identified parameter values and mean error of the reward function defined by (\ref{eq_reward_func_cart_coord_17}) for S1 }  \label{table_para_carte_sys_17_1}
\centering
\begin{tabular}{@{}ccccccc@{}}
\toprule
\multicolumn{7}{c}{\rule{0pt}{10pt} Mean error: 0.1288} \\ 
\hline
 $\theta_1$ & $\theta_2$ & $\theta_3$ & $\theta_4$ & $\theta_5$ & $\theta_6$ & $\theta_7$  \\ 
\midrule
0  &  0.0012  &  7.1e-04   & -0.0041   & 0.0053  & -4.2e-04 & -0.0099 \\
\midrule 
$\theta_8$ & $\theta_9$ & $\theta_{10}$ & $\theta_{11}$  & $ \theta_{12}$ & $\theta_{13}$ & $\theta_{14}$ \\
\midrule 
 2.7e-4 & 0.0018 & -0.0051 &  0.0135 & 0.0971 & 0.0963 & 0.1102  \\
\midrule 
$\theta_{15}$ & $\theta_{16}$ & $\theta_{17} $ & $\theta_{18} $  & $ \theta_{19}$  & $\theta_{20}$ &   \\
\midrule 
 -8.2e-05 & -0.0307 & 0.0167 & 0 & 0 & 0  \\
\bottomrule
\end{tabular}
\end{table}

\begin{table}
\caption{Identified parameter values and mean error of the reward function in (\ref{eq_6para_carte_coord_sys}) for S1}  \label{table_6para_carte_sys_lego_s1}
\centering
\begin{tabular}{@{}cccccc@{}}
\toprule
\multicolumn{6}{c}{\rule{0pt}{10pt} Mean error: 0.1611} \\\hline
 $\theta_{1}$ & $\theta_{2}$ & $\theta_{3}$ & $\theta_{4}$ & $\theta_{5}$ & $\theta_{6}$     \\
 \midrule
-0.0714 & 0.0842 & 0.0329 & 0.0914 & 0.2443 & 0.0275   \\
\bottomrule
\end{tabular}
\end{table}

\begin{table}
\caption{{Fit error of the reward function on the S1 dataset as a function of the number of retained basis functions}}
\label{table_ablation_num_param}
\centering
{
\begin{tabular}{@{}ccccc@{}}
\toprule
Parameters & 20 & 12 & 6 & 4 \\
\midrule
Max error & 0.7663 & 0.7445 & 0.7580 & 0.8127 \\
Mean error & 0.1288 & 0.1332 & 0.1611 & 0.1996 \\
\bottomrule
\end{tabular}
}
\end{table}

\begin{table}
\caption{Identified parameter values and mean error of the reward function in (\ref{eq_6para_carte_coord_sys}) for S2}  \label{table_6para_carte_sys_lego_s2}
\centering
\begin{tabular}{@{}cccccc@{}}
\toprule
\multicolumn{6}{c}{\rule{0pt}{10pt} Mean error: 0.1755} \\\hline
 $\theta_{1}$ & $\theta_{2}$ & $\theta_{3}$ & $\theta_{4}$ & $\theta_{5}$ & $\theta_{6}$     \\
 \midrule
0.0607 & 0.0829 & 0.0414 & 0.0910 & -0.2168 & 0.0277   \\
\bottomrule
\end{tabular}
\end{table}

\begin{table}
\caption{{Implementation settings used in the reported studies}}
\label{tab_implementation}
\centering
{
\begin{tabular}{@{}ll@{}}
\toprule
Setting & Value \\
\midrule
Operating domain $\Omega$ & radius $3$ (simulation), $0.37$\,m (experiments) \\
Dataset collection grid & $11 \times 21$ \\
Planning grid & $11 \times 12$ \\
Admissible actions $\mathcal{U}$ & $\{ \cdot, \uparrow, \downarrow, \leftarrow, \rightarrow \}$ \\
Particle number $N$ & $10\,000$ \\
Prior (simulation) & $\theta^i_0 \sim \mathcal{U}(-3,3)$ \\
Prior (experiments) & $2\times$ above and below identified values \\
Measurement noise & $\mathbb{E}[\mu_k]=0$, $\mathbb{E}[\mu_k^2]=0.25$ \\
Confidence threshold & $0.95$ (numerical), $0.9$ (high-fidelity) \\
Independent runs & $100$ \\
Object detector & YOLOv5-s \cite{jocher2022ultralyticsyolov5}, fine-tuned \\
Training set & $5184$ images, annotated in Isaac Sim \\
\bottomrule
\end{tabular}
}
\end{table}

\subsection{Numerical simulation}
\label{subsection_numerical_sim_aod_lego}
A numerical simulation was conducted using MATLAB to assess the effectiveness and performance of the proposed DCEE for identifying the next best camera viewpoint that generates the optimal trajectory. 
To further demonstrate the superior performance of the proposed DCEE algorithm, comparative studies are performed against competitive approaches, including MPC, entropy, and UCB methods, as described in Subsection \ref{sec_mpc_entropy}.

The ground truth values (\emph{i.e.}, identified values from datasets) of the unknown parameter $\theta$ for S1  are given in Table \ref{table_6para_carte_sys_lego_s1}.
To simulate a real environment, confidence level measurements, $ \mathcal{C}_k $, are subject to Gaussian noise, {with the noise statistics, the particle filter size and the prior on $\theta$ specified in Table \ref{tab_implementation}}.
The camera operates according to the model given by equation  (\ref{eq_agent_model_aod}), with admissible set $ u_k \in \mathcal{U} =  \{ \cdot,  \uparrow, \downarrow, \leftarrow, \rightarrow \} $, and the domain $  \Omega = \{ \mathrm{p}_k| \mathrm{p}_{x}^2+\mathrm{p}_{y}^2 + \mathrm{p}_{z}^2 = 9, ~ \mathrm{p}_{z} \geq 0 \} $.
At each movement, the camera advances to the next point along the hemisphere's surface, starting from an initial position of $ [-2.0175, -0.6555, 2.1213] $.
The average algorithm performance is evaluated using two key metrics: average convergence distance $\mathcal{{D}}= \| \mathrm{p}_k - \mathrm{p}^*_k \| $, where $ \mathrm{p}^*_k$ denotes the high-confidence viewpoint within the constrained domain, and the variance of parameter estimation $ \mathcal{P}= \mathrm{tr}\big({\mathrm{cov}}\{\theta_{k+1|k}\}\big)$, assessing the uncertainty in the estimated parameters.
For this scenario, set $ \mathrm{p}^*_{k} = [1.9635, 1.4266, 1.7634] $ within the defined domain.

\begin{remark}\label{rmk_conf_score_threshold}
If the measured confidence score exceeds a predefined threshold, such as $\mathcal{C}_k = 0.95$ in numerical simulations for S1, it indicates that the camera has reached the optimal viewpoint region, thereby completing the active object detection task.
This threshold-based criterion explains why, in some cases, the camera halts at different time steps.
In contrast, during online high-fidelity simulations, a slightly lower threshold (\emph{e.g.}, $ \mathcal{C}_{k} = 0.9 $) is utilized due to practical limitations, including degraded image quality due to object occlusion and environmental influences, and detector performance.  
From a practical perspective, achieving a satisfactory confidence score within the optimal region, while following the optimal trajectory, constitutes a reasonable and acceptable outcome for real-world applications.
The threshold is determined based on specific detection outcomes, for instance, by selecting viewpoints corresponding to the top $ 2 \% $ of the highest confidence scores.
\end{remark}

\begin{figure}
\centering
\input{Figures/path_s1_aod_dcee}
\caption{One representative trajectory of the camera for S1, illustrating movement from low-confidence regions on the negative $y$-axis toward higher-confidence areas on the positive side. 
Blue and red areas represent the obstacle and LEGO brick, respectively.
    }
\label{fig_view_trajec_estimatedpara_s1_v2.4}
\end{figure}
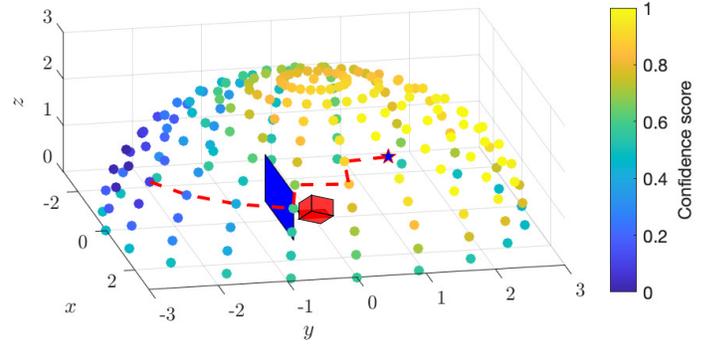

\begin{figure}
    \centering
    \includegraphics[width=0.5 \textwidth]{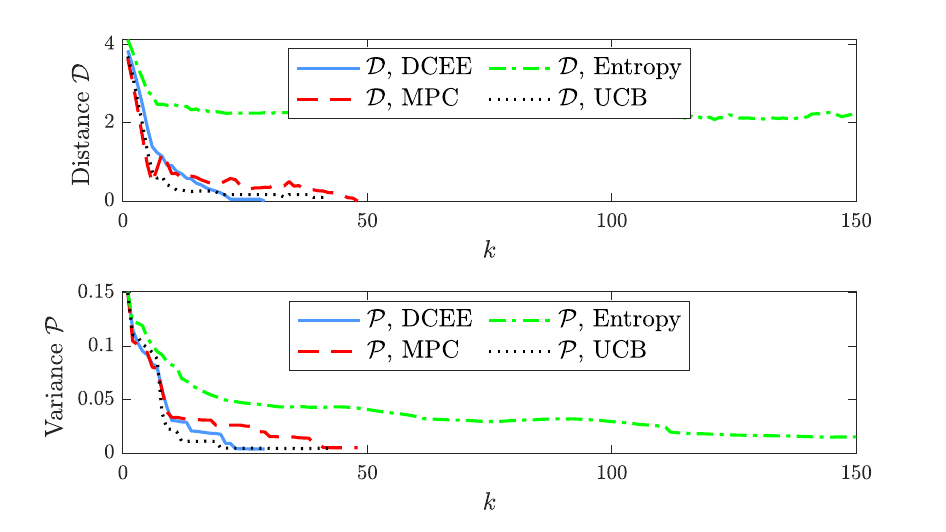}
    \caption{{Comparison of average convergence distance and parameter estimation variance for DCEE, MPC, entropy, and UCB (with $\beta=0.5$, selected from a series of parameter tuning trials) methods over $100$ independent MATLAB runs in S1.}
    }
    \label{fig_d_p_compare_dcee_entropy_mpc_ucb_lego_s1}
\end{figure}

Simulation results are presented in Figs. \ref{fig_view_trajec_estimatedpara_s1_v2.4}--\ref{fig_d_p_compare_dcee_entropy_mpc_ucb_lego_s1}.
Fig. \ref{fig_view_trajec_estimatedpara_s1_v2.4} illustrates one trajectory of the camera, which moves from regions of lower confidence on the negative side of the $y$-axis to areas of higher confidence on the opposite side. 
This movement demonstrates the effectiveness of the proposed DCEE algorithm for active object detection.
Comparative results among DCEE, MPC, and entropy methods are shown in Fig. \ref{fig_d_p_compare_dcee_entropy_mpc_ucb_lego_s1}, while the UCB baseline included in the same figure is discussed separately below.
The first sub-figure in Fig. \ref{fig_d_p_compare_dcee_entropy_mpc_ucb_lego_s1} presents the average convergence distance $\mathcal{{D}}$, and how it evolves for the three methods. 
Although MPC exhibits faster convergence during the initial phase, DCEE ultimately surpasses it in speed, whereas the entropy method demonstrates the slowest convergence throughout the entire process, along with the largest deviation.
This behavior is expected, as DCEE initially explores unknown objects and environments, resulting in slower convergence than MPC. 
However, once this exploratory phase is complete, DCEE achieves the fastest convergence speed.
The second sub-figure in Fig. \ref{fig_d_p_compare_dcee_entropy_mpc_ucb_lego_s1} illustrates how the variance of parameter estimation $\mathcal{P}$ updates over time among DCEE, MPC, and entropy methods.
Clearly, the variance of parameter convergence in DCEE exhibits the fastest convergence rate, coupled with the most efficient exploration. 
In contrast, MPC converges more slowly, while the entropy method displays the largest convergence error.
Table \ref{tab_s1} presents a quantitative comparative analysis of these attributes based on numerical simulation results.
The DCEE algorithm reaches the optimal viewpoint area within $28$ steps, while MPC requires approximately $ 46 $ steps, and the entropy method is the slowest, taking about $ 89 $ steps.

{%
To evaluate the sensitivity of DCEE to measurement noise, Table \ref{tab_noise_perturbation} reports the average convergence steps over $100$ independent runs for different noise variance levels.
DCEE converges at all tested levels, requiring $23$, $28$ and $30$ steps for $\mathbb{E}[\mu_k^2]=0.09$, $0.25$ and $0.36$, respectively.}

\begin{table}
\centering
\caption{{Convergence speed of DCEE under increasing measurement noise variance, realized as a perturbation of the identified reward model, over $100$ independent runs in S1}}
\label{tab_noise_perturbation}
{
\begin{tabular}{c|c}
\hline
{Noise level $\mathbb{E}[\mu_k^2]$} & \thead{Convergence steps [$\mathrm{k}$]}   \\
\hline
0.09 & 23  \\
0.25 & 28  \\
0.36 & 30  \\
\hline
\end{tabular}
}
\end{table}

These promising results will be further validated through high-fidelity simulations and real-world experiments.

{To further position the proposed method against recent active perception baselines, DCEE is compared with the UCB acquisition (\ref{eq_ucb}) of Subsection \ref{sec_mpc_entropy}, with $\beta=0.5$ selected from tested values through parameter tuning.
Fig. \ref{fig_d_p_compare_dcee_entropy_mpc_ucb_lego_s1} shows that UCB achieves comparable average convergence distance and parameter estimation variance to DCEE, converging in $42$ steps, compared with $28$, $46$ and $89$ steps for DCEE, MPC and entropy, respectively (Table \ref{tab_s1}).
Unlike DCEE, however, UCB requires the exploration coefficient $\beta$ to be specified in advance and possibly re-tuned for each new scenario.}

\subsection{Virtual environment simulation}
\label{sec_vir_env_sim_lego_s1}
To further validate the performance of the proposed DCEE algorithm in a high-fidelity environment built in Isaac Sim, the same settings used in the numerical simulations described in Subsection \ref{subsection_numerical_sim_aod_lego} are maintained, except that object confidence measurements are generated by the YOLOv5-s image processing model in \cite{jocher2022ultralyticsyolov5}, which processes captured RGB images.
The high-fidelity virtual environment in Isaac Sim integrates the camera and ROS2, facilitating communication among system components.
This integration enables efficient information exchange between key components, including image acquisition, the {YOLOv5-s model}, the DCEE algorithm, and camera movement control.
Notably, this online virtual simulation in Isaac Sim closely approximates real-world conditions \cite{kainova2023overview}, further reinforcing the practical applicability of the proposed approach.

\begin{figure}
    \centering
    \includegraphics[trim=1 1 0 2, clip,width=0.5 \textwidth]{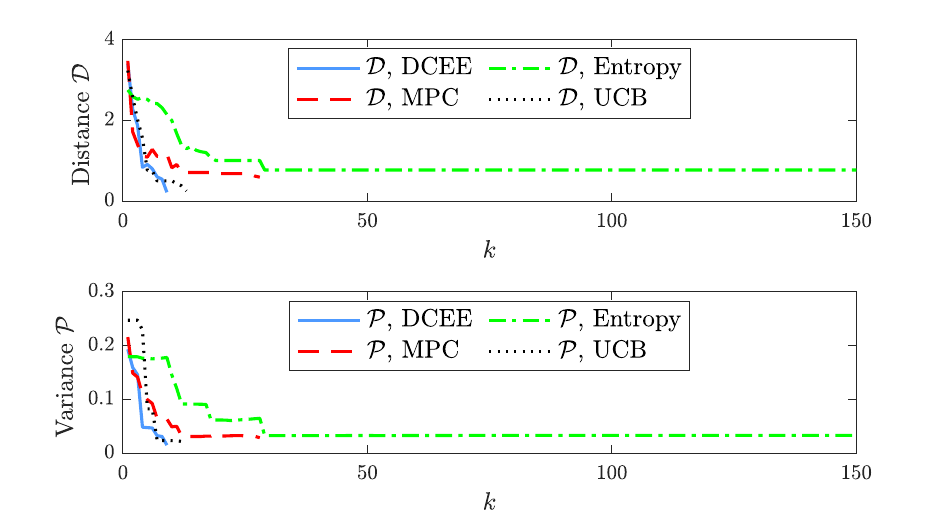}
    \caption{{Comparison of average convergence distance and estimation variance among DCEE, MPC, entropy, and UCB methods over $100$ independent runs in the virtual environment using Isaac Sim for S1.}
    }
    \label{fig_d_p_compare_lego_s1_ros2}
\end{figure}

The high-fidelity virtual simulation results are presented in Fig. \ref{fig_d_p_compare_lego_s1_ros2}, demonstrating the effectiveness and robustness of the DCEE algorithm. 
These results are consistent with the promising outcomes observed in Subsection \ref{subsection_numerical_sim_aod_lego}, where DCEE demonstrates faster convergence with smaller errors compared to the entropy, MPC, {and UCB} methods, both in terms of average convergence distance and estimation variance.
Note that in Fig. \ref{fig_d_p_compare_lego_s1_ros2}, the distance error converges to a small region around the origin because the camera movement stops when the detected confidence score reaches the predefined threshold, as mentioned in Remark \ref{rmk_conf_score_threshold}.
Table \ref{tab_s1} provides a comparative analysis of these attributes based on the online high-fidelity simulation results.
DCEE needs only $9$ steps to converge, while the entropy method takes $30$ steps, MPC takes $28$ steps, and {UCB, at its best coefficient, takes $13$ steps} in Table \ref{tab_s1}.

\begin{table}
\centering
\caption{Convergence performance of DCEE, MPC, entropy, and UCB methods in S1 simulations}
\label{tab_s1}
\begin{tabular}{c|c|c}
\hline
{Scenarios} & {Algorithms} & \thead{Convergence \\ steps [$\mathrm{k}$] }   \\
\hline
\multirow{4}{*}{S1 in MATLAB}  & DCEE & 28   \\
&  Entropy & 89   \\
&  MPC &  46   \\
&  {UCB} & {42}   \\
\hline
\centering
\multirow{4}{*}{S1 in Isaac Sim}  & DCEE & 9  \\
&  Entropy &  30    \\
&  MPC &  28    \\
&  {UCB} & {13}   \\
\hline
\end{tabular}
\end{table}

\section{Experiments for LEGO brick detection}

\begin{figure*}
    \centering
    \input{Figures/ros2_system_diagram.tex}
    \caption{Experimental system structure: UR5e robot arm (with end-mounted RealSense camera), host PC that includes  YOLOv5 detector and the DCEE-based GOCS, environment that includes the LEGO brick and obstacle (rectangular box), and ROS2 topics. }
    \label{fig:expt_struc}
\end{figure*}
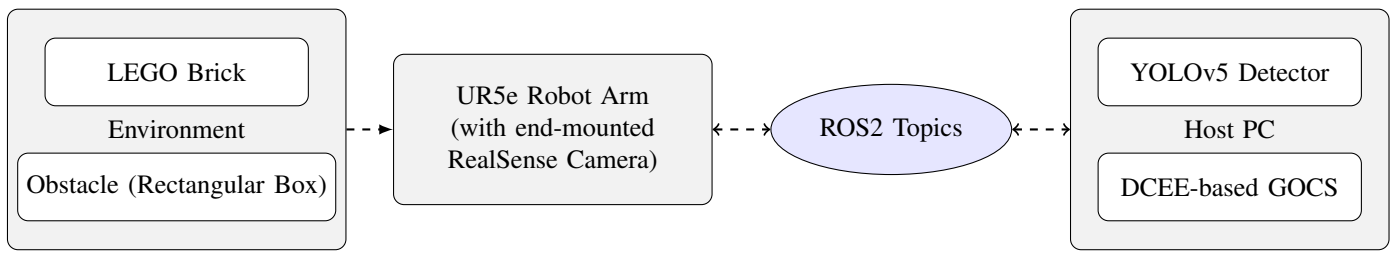 

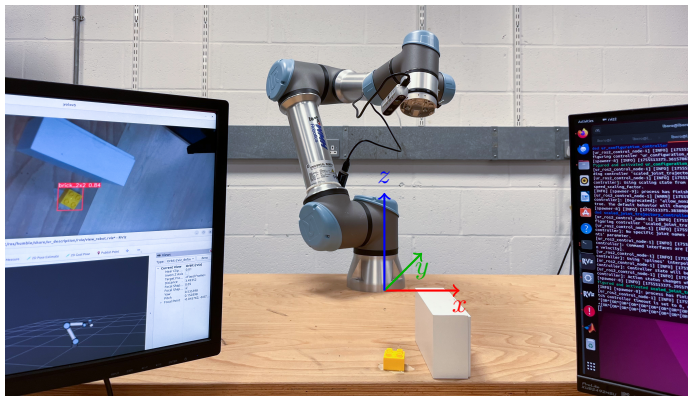
\begin{figure}
    \centering
    \input{Figures/expt_setting_xyz.tex}
\caption{A real-world environment is established, consisting of a yellow $2\times2$ LEGO brick, a white rectangular obstacle on the positive side of the $x$-axis, and a host PC. 
    This setup is designated as Scenario 3 (S3).}
    \label{fig_aod_lego_s3_expt}
\end{figure}

To further evaluate the effectiveness of the proposed DCEE algorithm in a real-world setting, we conducted LEGO brick detection experiments {in which the brick was placed at an arbitrary location within the operating region, and no prior information about its location was provided to the algorithm}.
The experimental setting integrates a UR5e robot arm (equipped with an end-mounted RealSense Camera), a host PC running the YOLOv5-s detector and the DCEE-based GOCS, and ROS2 topics that coordinate communication among system components, as illustrated in Fig. \ref{fig:expt_struc}.
This architecture enables efficient information exchange among critical modules, including image acquisition, the {YOLOv5-s model}, the DCEE algorithm, and the motion control of the UR5e robot.
ROS2 topics facilitate communication among the robot, camera, and host PC, while also enabling coordinated interactions between the YOLOv5 detector and the DCEE-based GOCS within the host PC.
In this study, the movement of the UR5e robot specifically refers to the motion of its end flange (\emph{i.e.}, the flange cap). 
A RealSense camera is rigidly mounted adjacent to the flange cap, as shown in Fig. \ref{fig_aod_lego_s3_expt}, ensuring that the camera’s position and orientation are directly coupled with the robot’s end-effector. 
Accordingly, precise control of the camera’s movement, including both translation and rotation, is achieved through the motion control of the robot’s flange cap.
The UR5e flange cap is actuated using the URScript \texttt{movej} command \cite{unirobots2009urscript}, which computes time-parameterized joint-space trajectories to ensure smooth, collision-free motion from the current pose to the DCEE-specified target pose with high kinematic precision. 
Object confidence scores are derived from the YOLOv5-s detector \cite{jocher2022ultralyticsyolov5}, which processes the acquired RGB images to provide reliable perception inputs.
The experimental results are available in the video access link: \url{https://youtu.be/FdFXst8uWxc?si=aAL23uZYahkuoJzX}.

\begin{figure}
    \centering
    \includegraphics[trim=15 5 20 10, clip,width=0.5 \textwidth]{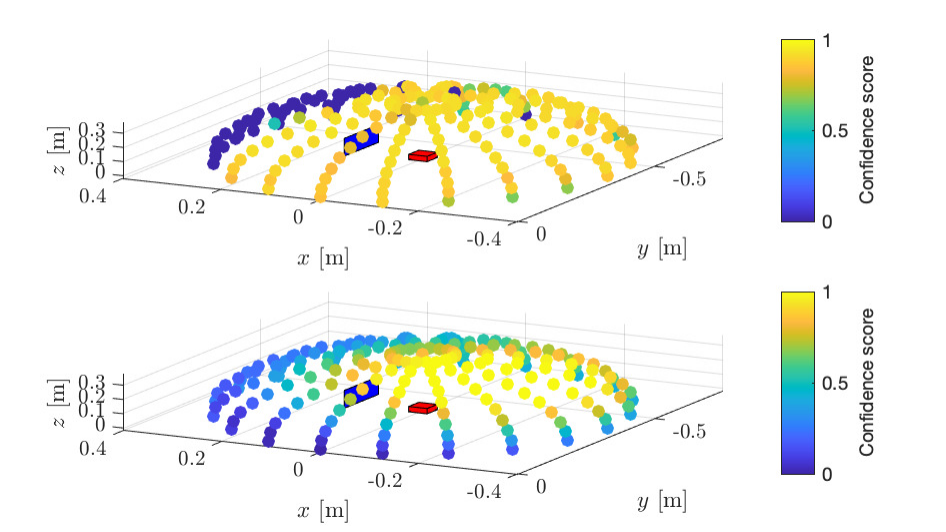}
    \caption{Collected datasets and the generated reward function with parameters listed in Table \ref{table_6para_carte_sys_lego_s1_expt} illustrate the confidence scores of the LEGO brick in S3 for the experiment setting, at various locations within a hemispherical domain, divided into an $11 \times 21 $ grid. The figure's blue and red areas represent the rectangular obstacle and LEGO brick for S3 given in Fig. \ref{fig_aod_lego_s3_expt}, respectively. }
    \label{fig_dataset_conf_scor_lego_s1_expt}
\end{figure}

When configuring the UR5e robot, the original Scenario 1 (S1) requires adjustment because the robot base is fixed at the origin, with the positive $y$-axis oriented toward the wall rather than the operational domain considered in this study (see Fig. \ref{fig_aod_lego_s3_expt}). 
To align the robot’s coordinate system with the experimental workspace, the environment is redefined as Scenario 3 (S3). 
This redefinition ensures consistency between the robot’s kinematic frame, the perception domain, and the task space, thereby enabling accurate implementation of the DCEE algorithm.
Building on the setup in Section \ref{sec_lego} for S1, S3 is defined as a hemispherical workspace $ \Omega = \{ \mathrm{p}_k| \mathrm{p}_{x}^2+(\mathrm{p}_{y}+0.3)^2 + \mathrm{p}_{z}^2 = 0.37^2, ~ \mathrm{p}_{z} \geq 0 \} $, centered at $(0, -0.3, 0)$ and discretized into an $11 \times 21$ grid, as shown in Fig. \ref{fig_aod_lego_s3_expt}.
{The radius of the domain follows from the reach and the collision-free workspace of the UR5e robot.}
In this setting, the LEGO brick is placed at $(0, -0.3, 0)$ on the negative $y$-axis, and the obstacle is located on the positive $x$-axis. 
Image datasets are then collected within this domain, following Section \ref{subsec_reward_funs}, to estimate the reward function parameters.
The collected images are processed by the well-trained YOLOv5-s model \cite{jocher2022ultralyticsyolov5}.
After obtaining the confidence scores for the LEGO brick,
the reward function defined by (\ref{eq_reward_func_cart_coord_17}) is identified following the same procedure as in S1.
The resulting outcomes are shown in Fig. \ref{fig_dataset_conf_scor_lego_s1_expt}, and the corresponding identified parameters are summarized in Table \ref{table_6para_carte_sys_lego_s1_expt}.

\begin{table}
\caption{Identified parameter values and mean error of the reward function in (\ref{eq_6para_carte_coord_sys}) for S3 experiments}  
\label{table_6para_carte_sys_lego_s1_expt}
\centering
\begin{tabular}{@{}cccccc@{}}
\toprule
\multicolumn{6}{c}{\rule{0pt}{10pt} Mean error: 0.2597} \\\hline
 $\theta_{1}$ & $\theta_{2}$ & $\theta_{3}$ & $\theta_{4}$ & $\theta_{5}$ & $\theta_{6}$     \\
 \midrule
26.447 & 1.6016 & 0.7485 & 11.4793 & -3.6871 & -6.3225   \\
\bottomrule
\end{tabular}
\end{table}

\begin{figure}[!t]
    \centering 
    \subfloat[$t = 6\,\mathrm{s}$]{\includegraphics[width=0.48\columnwidth]{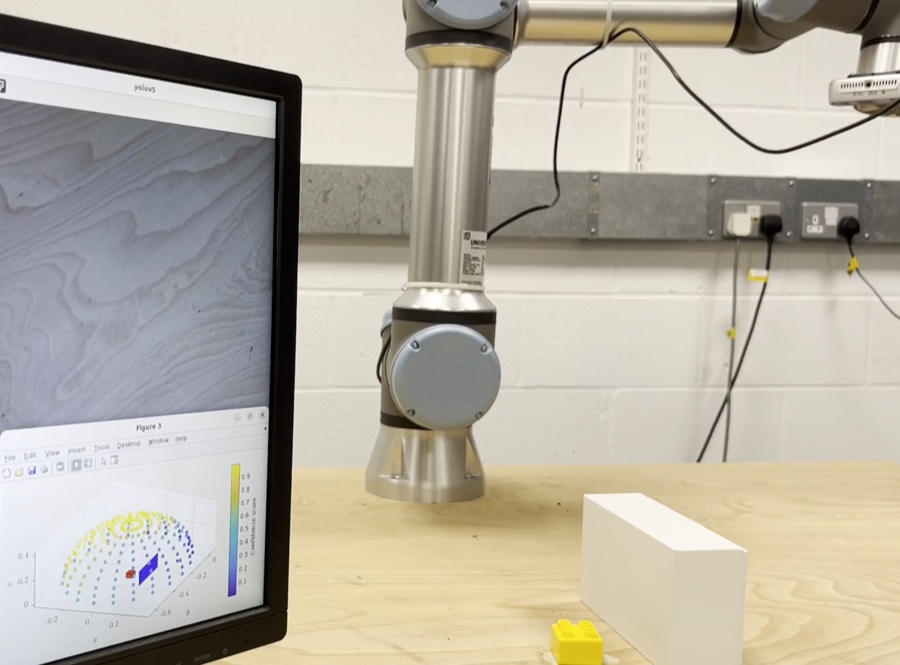}%
    \label{fig:1a}}
    \hfil
    \subfloat[$t = 9.5\,\mathrm{s}$]{\includegraphics[width=0.48\columnwidth]{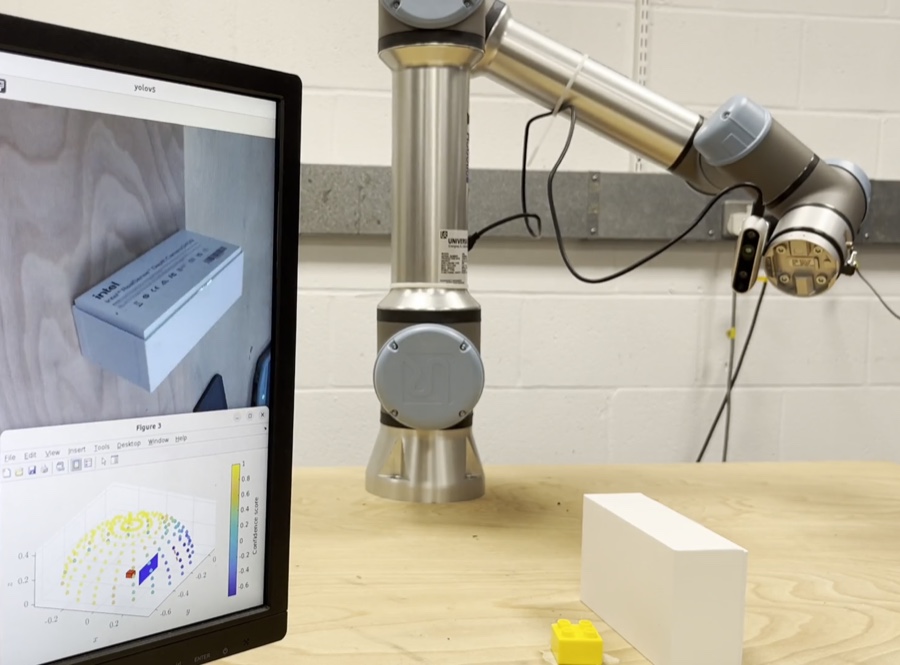}%
    \label{fig:1b}}
    \\[2mm]
    \subfloat[$t = 36\,\mathrm{s}$]{\includegraphics[width=0.48\columnwidth]{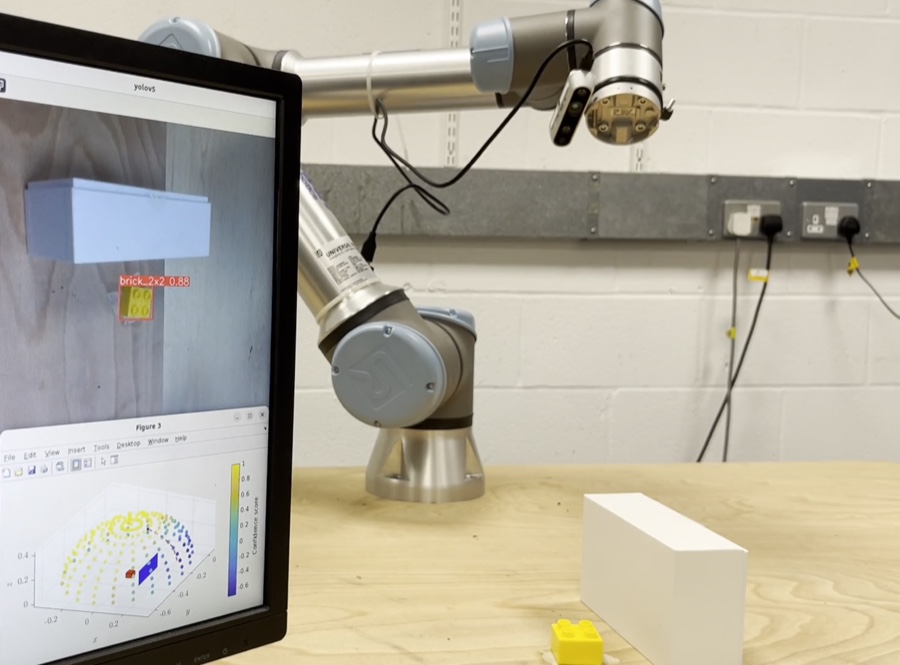}%
    \label{fig:1c}}
    \hfil
    \subfloat[$t = 42\,\mathrm{s}$]{\includegraphics[width=0.48\columnwidth]{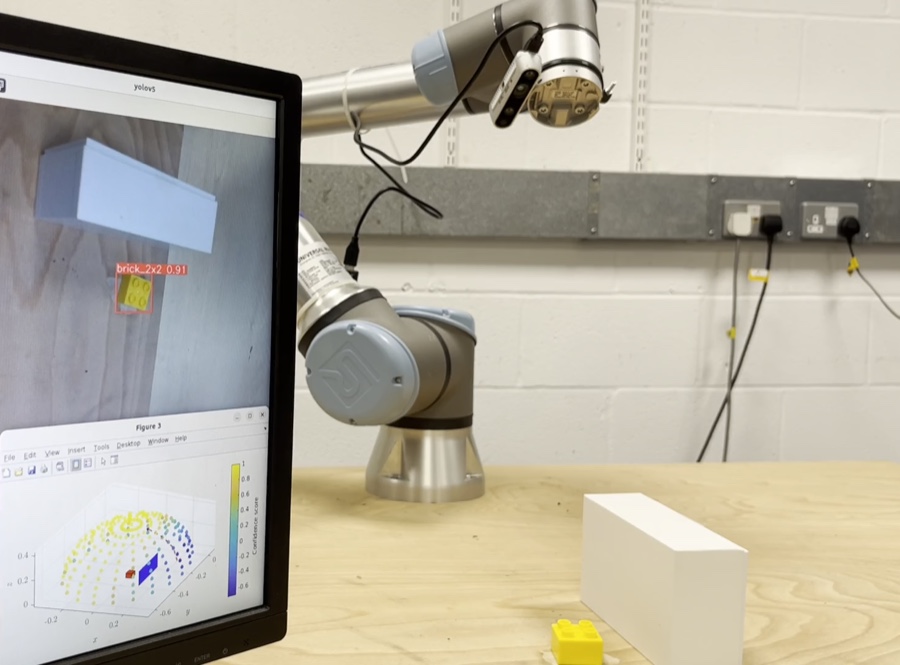}%
    \label{fig:1d}}
    \caption{Frame-by-frame plots of one experiment to show selected camera viewpoint movement sequences from the initial state to the terminal state: (a–b) the LEGO is occluded, (c) the LEGO is detected with confidence 0.88, and (d) the LEGO is detected with confidence 0.91. }
    \label{fig:frame-by-frame-plots-expts}
\end{figure}

\begin{figure}
\centering
\includegraphics[trim=5 5 20 10, clip,width=0.5 \textwidth]{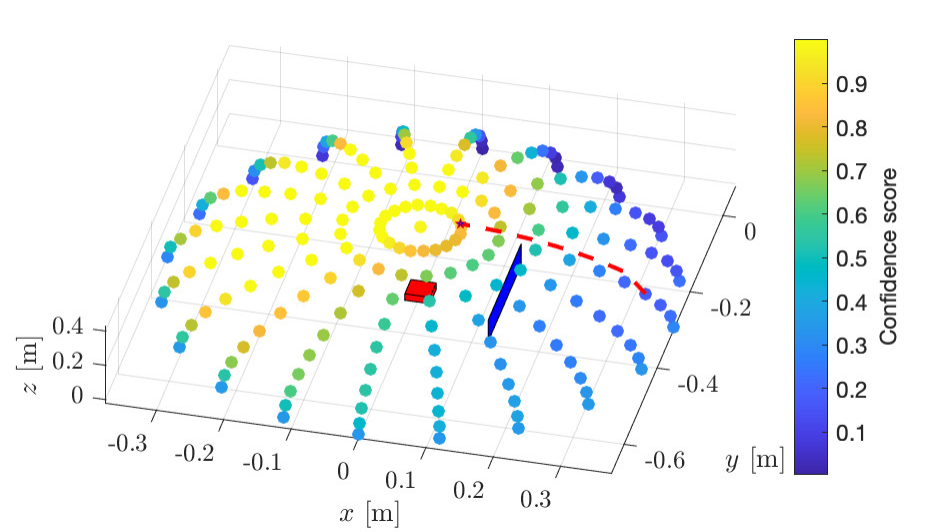}
\caption{
One representative trajectory of the camera for S3 in the experiments, showing movement from low-confidence regions on the positive $x$-axis toward higher-confidence areas. 
Blue and red regions indicate the obstacle and LEGO brick, respectively.
}
\label{fig_view_trajec_lego_s3_expt}
\end{figure}

\begin{figure}
    \centering
    \includegraphics[trim=20 2 20 12, clip,width=0.5 \textwidth]{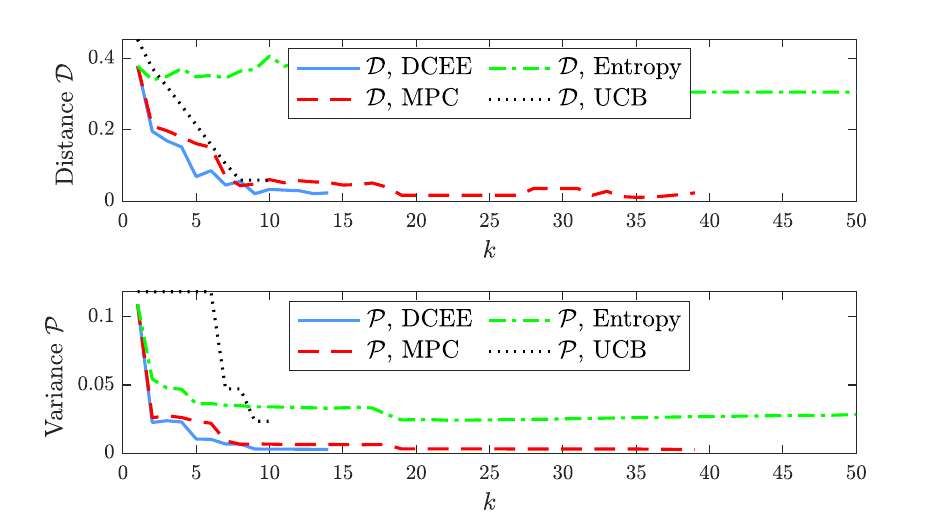}
    \caption{ {Comparison of average convergence distance and parameter estimation variance for DCEE, MPC, entropy, and UCB methods over $100$ independent runs in S3 experiments, as shown in Fig. \ref{fig_aod_lego_s3_expt}.}
    }
    \label{fig_d_p_compare_ros2_lego_s3_expt}
\end{figure}

Experiments in S3 are conducted to evaluate the performance of the proposed DCEE algorithm, starting from an initial position of $[0.3341, -0.3, 0.1702]$, selected to represent a typical starting pose from which the object of interest is not visible.
The target position $\mathrm{p}^*_{k} = [0.0937, -0.5885, 0.2204]$ lies within the workspace, offering a good viewpoint and a meaningful task for evaluating the algorithm’s performance. 
The experiments evaluate the overall performance of the DCEE algorithm in comparison to MPC, entropy, {and UCB} methods, all runs starting from the same initial state, {with the number of runs and the prior on the parameters as listed in Table \ref{tab_implementation}}.

The experimental results are presented in Figs. \ref{fig:frame-by-frame-plots-expts}–\ref{fig_d_p_compare_ros2_lego_s3_expt}.
Frame-by-frame plots of one experiment are shown in Fig. \ref{fig:frame-by-frame-plots-expts}, illustrating selected camera viewpoint sequences from the initial to the final state, where in subfigure \ref{fig:1d} the LEGO is detected with a confidence score of 0.91, highlighted by the red bounding box and its label.
Fig. \ref{fig_view_trajec_lego_s3_expt} illustrates a representative trajectory of the camera under the DCEE method, demonstrating its ability to move from a low-confidence region on the positive side of the $x$-axis toward higher-confidence areas across the workspace.
Fig. \ref{fig_d_p_compare_ros2_lego_s3_expt} provides a comparative analysis of DCEE, MPC, entropy, {and UCB} methods, highlighting the superior performance of DCEE in terms of both average convergence distance and parameter estimation variance.
The first sub-figure in Fig. \ref{fig_d_p_compare_ros2_lego_s3_expt} shows the evolution of the average convergence distance $\mathcal{{D}}$ for the {four} methods.
Although MPC initially converges similarly to DCEE and faster than the entropy method, DCEE ultimately achieves faster convergence, while the entropy method demonstrates the slowest convergence with substantial deviation.
{UCB converges more slowly than DCEE during the initial phase but reaches a comparable final distance within $10$ steps.}
The second sub-figure in Fig. \ref{fig_d_p_compare_ros2_lego_s3_expt} presents the variance of parameter estimation $\mathcal{P}$, further highlighting DCEE's faster convergence compared to MPC, entropy, {and UCB}.
Additionally, Table \ref{tab_s3_expt} provides a comprehensive numerical comparison of the performance metrics from the experimental results.
DCEE achieves convergence in $9$ steps, whereas {UCB, MPC, and entropy require $10$, $20$ and $23$} steps, respectively.
{The computational cost of DCEE scales with the number of admissible actions and the particle number $N$. On the experimental system (Intel Core i5 CPU, Ubuntu 20.04, ROS2 Humble), the mean reward computation time is $3.08\,\mathrm{s}$ per decision, while the mean total iteration time, including robot operation and belief updating, is $4.39\,\mathrm{s}$.
With an average of $9$ iterations (Table \ref{tab_s3_expt}), 
a complete run takes approximately $40\,\mathrm{s}$.}
The superior performance of DCEE stems from its active learning capability, achieved by incorporating the variance of posterior parameter estimation into the cost function. This design ensures that the control action contributes not only to camera movement (\emph{i.e.}, exploitation) but also to parameter estimation (\emph{i.e.}, exploration).

\begin{table}
\centering
\caption{{Convergence performance of DCEE, MPC, entropy, and UCB methods in S3 experiments}}
\label{tab_s3_expt}
\begin{tabular}{c|c|c}
\hline
{Scenarios} & {Algorithms} & \thead{Convergence \\ steps [$\mathrm{k}$] }   \\
\hline
\centering
\multirow{4}{*}{S3 in experiments}  & DCEE & 9  \\
&  Entropy &  23    \\
&  MPC &  20    \\
&  {UCB} & {10}   \\
\hline
\end{tabular}
\end{table}

\section{Application to peg detection with two obstacles}
\label{sec_peg_s4}

{To evaluate generalization beyond the LEGO brick, Scenario 4 (S4) applies the same DCEE framework to an industrial peg observed under two obstacles against an industrial textured background, evaluated in the high-fidelity virtual environment built in Isaac Sim.
The operating domain, its discretization, the particle filter settings, and the six-parameter reward-model structure are unchanged from Subsection \ref{subsection_numerical_sim_aod_lego}; only the object detector is fine-tuned on a peg dataset, following the same procedure as for the LEGO brick.
Following the protocol used throughout, DCEE is compared against MPC, entropy, and the UCB baseline of Subsection \ref{sec_mpc_entropy} in average convergence distance, estimation variance, and convergence steps.
The scenario is illustrated in Fig. \ref{fig_peg_s4_setup}.
The high-fidelity simulation results in Isaac Sim for S4 are available in the video access link: \url{https://youtu.be/loCzEHEvhfA}.}

\begin{figure}
    \centering
    \input{Figures/peg_s4/Demo_s4_peg_setup}
    \caption{{S4 scenario setting in Isaac Sim: the peg (gray) is placed away from the domain origin, with two cardboard-box obstacles against an industrial textured background. The origin is marked by a red cone, and the $x$, $y$, $z$ axes are shown as red, green, and blue arrows, respectively.}}
    \label{fig_peg_s4_setup}
\end{figure}

\begin{figure}
    \centering
    \includegraphics[trim=32 5 22 10, clip, width=0.5\textwidth]{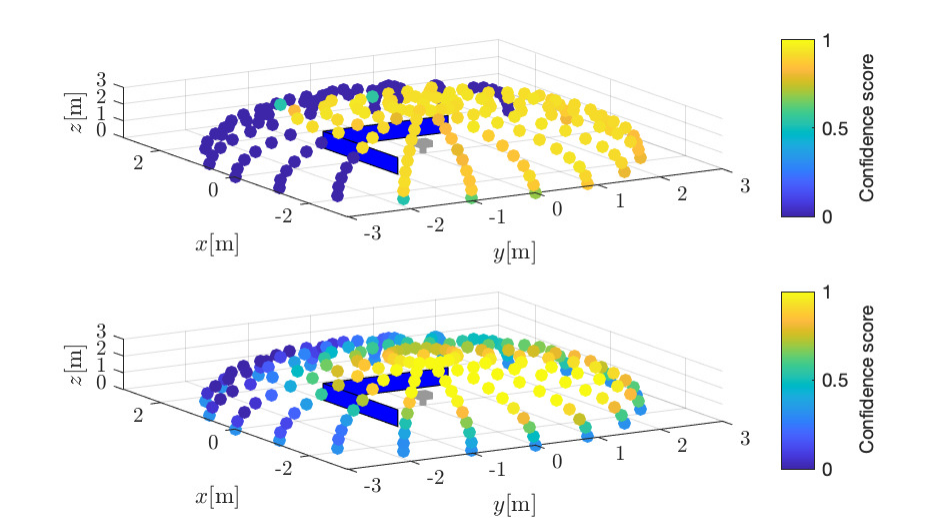}
    \caption{{Collected datasets (top) and the generated reward function (bottom) illustrate the confidence scores of the peg detected by the object detector fine-tuned for S4, acquired within the same hemispherical domain, divided into an $11\times21$ grid. The gray area represents the peg, and the blue area denotes two obstacles.}}
    \label{fig_datasets_mapping_peg_s4}
\end{figure}

{The collected dataset and the identified reward-model parameters for S4 follow the same procedure as S1 (Subsection \ref{subsec_reward_funs}), shown in Fig. \ref{fig_datasets_mapping_peg_s4} and Table \ref{table_6para_carte_sys_peg_s4}.}

\begin{table}
\caption{{Identified parameter values and mean error of the reward function in (\ref{eq_6para_carte_coord_sys}) for S4}}
\label{table_6para_carte_sys_peg_s4}
\centering
{
\begin{tabular}{@{}cccccc@{}}
\toprule
\multicolumn{6}{c}{\rule{0pt}{10pt} Mean error: 0.2046} \\\hline
 $\theta_{1}$ & $\theta_{2}$ & $\theta_{3}$ & $\theta_{4}$ & $\theta_{5}$ & $\theta_{6}$     \\
 \midrule
-0.0803 & 0.0355 & 0.0359 & 0.0974 & 0.2135 & -0.1036   \\
\bottomrule
\end{tabular}
}
\end{table}

\begin{figure}
    \centering
    \includegraphics[trim=5 5 20 10, clip, width=0.5\textwidth]{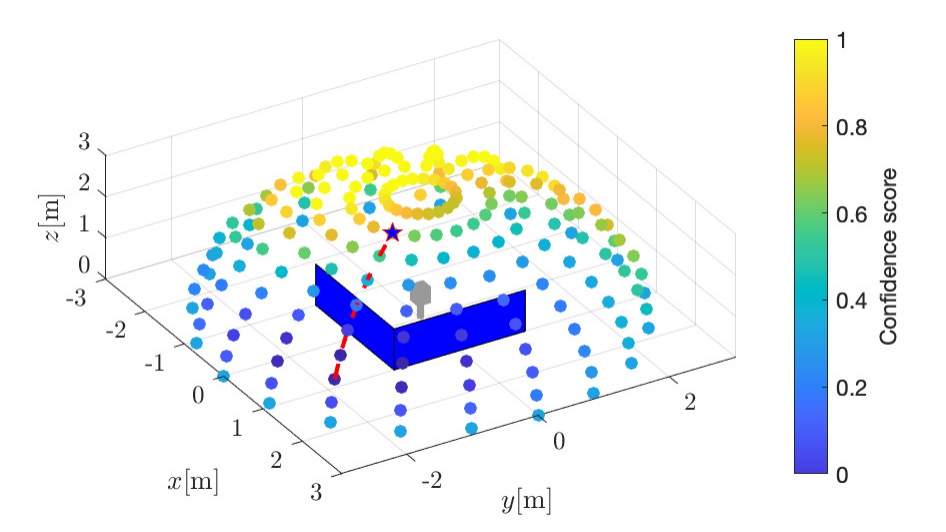}
    \caption{{One representative trajectory of the camera for S4 using the DCEE algorithm, moving from a low-confidence region behind the two obstacles toward higher-confidence areas. The blue area represents the two obstacles, and the gray icon marks the peg.}}
    \label{fig_view_trajec_peg_s4}
\end{figure}

{One representative trajectory of the camera for S4 using the DCEE algorithm is shown in Fig. \ref{fig_view_trajec_peg_s4}, illustrating movement from a low-confidence region behind the two obstacles toward higher-confidence areas, demonstrating the effectiveness of DCEE for active object detection on the peg.}

\begin{figure}
    \centering
    \includegraphics[trim=.5 1 .5 2, clip, width=0.5\textwidth]{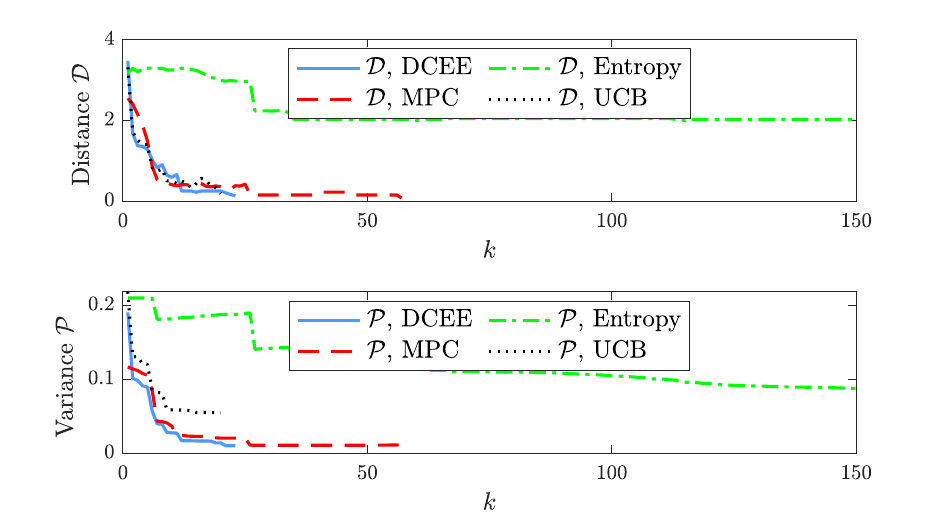}
    \caption{{Comparison of average convergence distance and estimation variance among DCEE, MPC, entropy, and UCB methods over $100$ independent runs for S4 (peg detection with two obstacles) in the virtual environment using Isaac Sim.}}
    \label{fig_d_p_compare_peg_s4}
\end{figure}

{Results are shown in Fig. \ref{fig_d_p_compare_peg_s4} and Table \ref{tab_s4}: DCEE converges fastest, requiring $12$ steps, followed by UCB with its best-performing coefficient ($20$ steps), MPC ($27$ steps), and entropy ($35$ steps). This confirms that the performance ordering observed in LEGO brick detection is preserved despite changes in the object and occlusions.}

\begin{table}
\centering
\caption{{Convergence performance of DCEE, MPC, entropy, and UCB methods in S4 simulations}}
\label{tab_s4}
{
\begin{tabular}{c|c|c}
\hline
{Scenarios} & {Algorithms} & \thead{Convergence \\ steps [$\mathrm{k}$]} \\
\hline
\multirow{4}{*}{S4 in Isaac Sim} & DCEE & 12 \\
& Entropy & 35 \\
& MPC & 27 \\
& UCB & 20 \\
\hline
\end{tabular}
}
\end{table}

\section{Conclusions}
This paper presents the development of the DCEE algorithm within goal-oriented control systems for active object detection, utilizing an exploration-exploitation balanced cost function based on visual information from a camera. 
The proposed DCEE algorithm actively explores optimal viewpoints by introducing variance-based uncertainty estimation while addressing challenges related to parameter estimation under uncertain measurements.
The designed linear regression-based reward function requires only six parameters to encode knowledge about variation in confidence scores as a function of viewpoint position within a domain. 
{The applicability of this function has been demonstrated across the scenarios considered.} 
The performance of the DCEE algorithm is validated through numerical simulations, high-fidelity virtual simulations, and real-world experiments, demonstrating its ability to effectively balance exploration and exploitation.
The algorithm's superior performance is further demonstrated through comparisons with existing methods (\emph{i.e.}, MPC, entropy, {and UCB}) across various scenarios, including LEGO bricks in different locations {and an industrial peg under occlusion}, achieving the expected results.
In this study, the camera is restricted to a hemispherical surface with a fixed orientation toward a designated reference point. 
These constraints streamline the analysis by minimizing extraneous complexity, thereby facilitating a focused evaluation of the algorithm's fundamental robustness. 
{These simplifications are appropriate when the object lies within a known region of interest that can be viewed from a surrounding hemisphere, as in the inspection and assembly settings considered here.}
{Further research will consider multiple objects of interest and occlusions that give rise to multiple locally optimal viewpoints, thereby relaxing the unique optimum assumption discussed in Remark \ref{rmk_scope_concavity}.}
{Further directions include modeling detection success and failure as a probabilistic event coupled to the reward, accommodating more generalized camera movements and orientations, developing computationally efficient action-selection strategies for continuous action spaces, accounting for dynamic occlusions and cluttered scenes, and explicitly incorporating safety constraints into the optimization framework.}

\bibliography{active_object_detection.bib}
\bibliographystyle{IEEEtran}

\newpage

\vfill

\end{document}

%% file: Figures/Demo_AOD_DCEE.tex
\tikzset{every picture/.style={line width=0.75pt}} %

\begin{tikzpicture}
\node at (0, 0) { \includegraphics[width=0.5\textwidth]{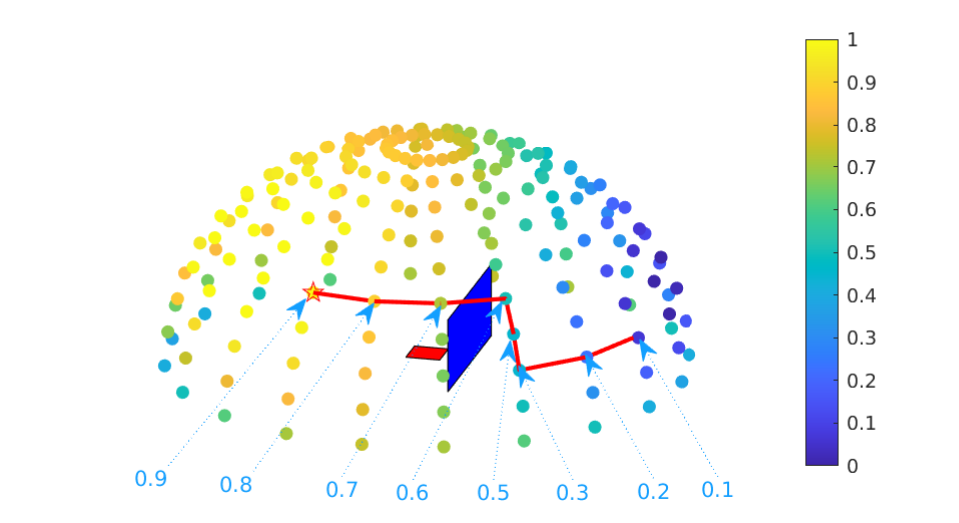} };
\node at (-3.2, -0.6) { 
\begin{tikzpicture}
 \robotArm[
config={q1=100,q2=-82},geometry={a1=0.8,a2=1},
end effector/.cd,
draw link/.code={
\draw[line cap=round, double=lightgray,
double distance=4mm]
(0,0) to[left] (2, 0.8);  },
draw end effector/.code={
\path[link style]
(1.8, 0.8) arc (180:0:0.2) -- (2.1, 0.4)
-- (1.9, 0.4) -- cycle;
\path[link style]
(2.05,0.4) rectangle (1.95,0.0);
\fill (2,0) circle (0.05);}
]{2}
\end{tikzpicture}
    };
    \node at (-1.2, -0.35){
    \begin{tikzpicture}[scale=0.3]
    \def\rotationAngle{-10} %
    
    \begin{scope}[rotate=\rotationAngle]
        
        \draw[thick] (-0.5,-0.3) rectangle (0.5,0.3); 
        \draw[thick] (0.5,0) circle (0.1); %
        
        \draw[thick, dashed] (0.5,0) -- (4,2);  %
        \draw[thick, dashed] (0.5,0) -- (4,-2); %
        
        \fill[blue!20, opacity=0.3] (0.5,0) -- (4,-2) arc [start angle=-32.57, end angle=26.57, radius=4cm] -- cycle;

    \end{scope}

\end{tikzpicture}
    };

    \node at (-0.5, -0.84){
\begin{tikzpicture}
    \begin{axis}[
        view={120}{30}, %
        hide axis, %
        xmin=0, xmax=3,
        ymin=0, ymax=3,
        zmin=0, zmax=3
    ]
        \def\a{0.2}
        
        \addplot3 [
            fill=red, opacity=0.8
        ] coordinates {
            (0,0,0) (0+\a,0,0) (0+\a,0+\a,0) (0,0+\a,0) (0,0,0)
        };
        \addplot3 [
            fill=red, opacity=0.8
        ] coordinates {
            (0,0,0) (0,0+\a,0) (0,0+\a,0+\a) (0,0,0+\a) (0,0,0)
        };
        \addplot3 [
            fill=red, opacity=0.8
        ] coordinates {
            (0,0,0) (0+\a,0,0) (0+\a,0,0+\a) (0,0,0+\a) (0,0,0)
        };
    \end{axis}
\end{tikzpicture}

    };
\end{tikzpicture}

%% file: Figures/AOD_flow_chart.tex
\tikzset{every picture/.style={line width=0.75pt}} %

\begin{tikzpicture}[x=0.75pt,y=0.75pt,yscale=-1,xscale=1]

\draw  [fill={rgb, 255:red, 255; green, 255; blue, 255 }  ,fill opacity=1 ] (95.75,82) -- (207.25,82) .. controls (209.06,82) and (210.53,83.47) .. (210.53,85.28) -- (210.53,98.72) .. controls (210.53,100.53) and (209.06,102) .. (207.25,102) -- (95.75,102) .. controls (93.94,102) and (92.47,100.53) .. (92.47,98.72) -- (92.47,85.28) .. controls (92.47,83.47) and (93.94,82) .. (95.75,82) -- cycle ;
\draw  [fill={rgb, 255:red, 255; green, 255; blue, 255 }  ,fill opacity=1 ] (95.75,126.4) -- (252.75,126.4) .. controls (254.56,126.4) and (256.03,127.87) .. (256.03,129.68) -- (256.03,143.12) .. controls (256.03,144.93) and (254.56,146.4) .. (252.75,146.4) -- (95.75,146.4) .. controls (93.94,146.4) and (92.47,144.93) .. (92.47,143.12) -- (92.47,129.68) .. controls (92.47,127.87) and (93.94,126.4) .. (95.75,126.4) -- cycle ;
\draw  [fill={rgb, 255:red, 255; green, 255; blue, 255 }  ,fill opacity=1 ] (96.25,173) -- (262.75,173) .. controls (264.56,173) and (266.03,174.47) .. (266.03,176.28) -- (266.03,189.72) .. controls (266.03,191.53) and (264.56,193) .. (262.75,193) -- (96.25,193) .. controls (94.44,193) and (92.97,191.53) .. (92.97,189.72) -- (92.97,176.28) .. controls (92.97,174.47) and (94.44,173) .. (96.25,173) -- cycle ;
\draw  [fill={rgb, 255:red, 255; green, 255; blue, 255 }  ,fill opacity=1 ] (229.75,82) -- (335.25,82) .. controls (337.06,82) and (338.53,83.47) .. (338.53,85.28) -- (338.53,98.72) .. controls (338.53,100.53) and (337.06,102) .. (335.25,102) -- (229.75,102) .. controls (227.94,102) and (226.47,100.53) .. (226.47,98.72) -- (226.47,85.28) .. controls (226.47,83.47) and (227.94,82) .. (229.75,82) -- cycle ;
\draw   (361,84.14) .. controls (361,66) and (375.71,51.29) .. (393.86,51.29) -- (614.39,51.29) .. controls (632.54,51.29) and (647.25,66) .. (647.25,84.14) -- (647.25,182.71) .. controls (647.25,200.86) and (632.54,215.57) .. (614.39,215.57) -- (393.86,215.57) .. controls (375.71,215.57) and (361,200.86) .. (361,182.71) -- cycle ;
\draw   (391.75,59.27) -- (610.25,59.27) .. controls (612.06,59.27) and (613.53,60.74) .. (613.53,62.55) -- (613.53,75.99) .. controls (613.53,77.8) and (612.06,79.27) .. (610.25,79.27) -- (391.75,79.27) .. controls (389.94,79.27) and (388.47,77.8) .. (388.47,75.99) -- (388.47,62.55) .. controls (388.47,60.74) and (389.94,59.27) .. (391.75,59.27) -- cycle ;
\draw  [fill={rgb, 255:red, 255; green, 255; blue, 255 }  ,fill opacity=1 ] (392.25,96.36) -- (609.25,96.36) .. controls (611.06,96.36) and (612.53,97.83) .. (612.53,99.64) -- (612.53,113.08) .. controls (612.53,114.89) and (611.06,116.36) .. (609.25,116.36) -- (392.25,116.36) .. controls (390.44,116.36) and (388.97,114.89) .. (388.97,113.08) -- (388.97,99.64) .. controls (388.97,97.83) and (390.44,96.36) .. (392.25,96.36) -- cycle ;
\draw  [fill={rgb, 255:red, 255; green, 255; blue, 255 }  ,fill opacity=1 ] (392.25,143.75) -- (609.75,143.75) .. controls (611.56,143.75) and (613.03,145.22) .. (613.03,147.03) -- (613.03,160.47) .. controls (613.03,162.28) and (611.56,163.75) .. (609.75,163.75) -- (392.25,163.75) .. controls (390.44,163.75) and (388.97,162.28) .. (388.97,160.47) -- (388.97,147.03) .. controls (388.97,145.22) and (390.44,143.75) .. (392.25,143.75) -- cycle ;
\draw  [fill={rgb, 255:red, 255; green, 255; blue, 255 }  ,fill opacity=1 ] (370.75,179) -- (639.25,179) .. controls (641.06,179) and (642.53,180.47) .. (642.53,182.28) -- (642.53,195.72) .. controls (642.53,197.53) and (641.06,199) .. (639.25,199) -- (370.75,199) .. controls (368.94,199) and (367.47,197.53) .. (367.47,195.72) -- (367.47,182.28) .. controls (367.47,180.47) and (368.94,179) .. (370.75,179) -- cycle ;
\draw  [fill={rgb, 255:red, 255; green, 255; blue, 255 }  ,fill opacity=1 ] (95.5,213.75) -- (232.17,213.75) .. controls (233.98,213.75) and (235.45,215.22) .. (235.45,217.03) -- (235.45,230.47) .. controls (235.45,232.28) and (233.98,233.75) .. (232.17,233.75) -- (95.5,233.75) .. controls (93.69,233.75) and (92.22,232.28) .. (92.22,230.47) -- (92.22,217.03) .. controls (92.22,215.22) and (93.69,213.75) .. (95.5,213.75) -- cycle ;
\draw    (181.83,102) -- (181.83,124.33) ;
\draw [shift={(181.83,126.33)}, rotate = 270] [color={rgb, 255:red, 0; green, 0; blue, 0 }  ][line width=0.75]    (10.93,-3.29) .. controls (6.95,-1.4) and (3.31,-0.3) .. (0,0) .. controls (3.31,0.3) and (6.95,1.4) .. (10.93,3.29)   ;
\draw    (260,102) -- (260,171) ;
\draw [shift={(260,173)}, rotate = 270.6] [color={rgb, 255:red, 0; green, 0; blue, 0 }  ][line width=0.75]    (10.93,-3.29) .. controls (6.95,-1.4) and (3.31,-0.3) .. (0,0) .. controls (3.31,0.3) and (6.95,1.4) .. (10.93,3.29)   ;
\draw    (338.17,92) -- (359,92.3) ;
\draw [shift={(361,92.33)}, rotate = 180.74] [color={rgb, 255:red, 0; green, 0; blue, 0 }  ][line width=0.75]    (10.93,-3.29) .. controls (6.95,-1.4) and (3.31,-0.3) .. (0,0) .. controls (3.31,0.3) and (6.95,1.4) .. (10.93,3.29)   ;
\draw    (265.75,183) -- (358.91,182.91) ;
\draw [shift={(360.91,182.91)}, rotate = 179.95] [color={rgb, 255:red, 0; green, 0; blue, 0 }  ][line width=0.75]    (10.93,-3.29) .. controls (6.95,-1.4) and (3.31,-0.3) .. (0,0) .. controls (3.31,0.3) and (6.95,1.4) .. (10.93,3.29)   ;
\draw    (502.36,79.27) -- (502.34,94.36) ;
\draw [shift={(502.33,96.36)}, rotate = 270.09] [color={rgb, 255:red, 0; green, 0; blue, 0 }  ][line width=0.75]    (10.93,-3.29) .. controls (6.95,-1.4) and (3.31,-0.3) .. (0,0) .. controls (3.31,0.3) and (6.95,1.4) .. (10.93,3.29)   ;
\draw    (501.09,116.36) -- (501.12,141.75) ;
\draw [shift={(501.13,143.75)}, rotate = 269.89] [color={rgb, 255:red, 0; green, 0; blue, 0 }  ][line width=0.75]    (10.93,-3.29) .. controls (6.95,-1.4) and (3.31,-0.3) .. (0,0) .. controls (3.31,0.3) and (6.95,1.4) .. (10.93,3.29)   ;
\draw    (500.76,163.75) -- (500.73,177) ;
\draw [shift={(500.73,179)}, rotate = 270.12] [color={rgb, 255:red, 0; green, 0; blue, 0 }  ][line width=0.75]    (10.93,-3.29) .. controls (6.95,-1.4) and (3.31,-0.3) .. (0,0) .. controls (3.31,0.3) and (6.95,1.4) .. (10.93,3.29)   ;
\draw    (505,215) -- (505,221.75) ;
\draw    (505,221.75) -- (350.4,223.75) ;
\draw [shift={(350.4,223.75)}, rotate = 359.42] [color={rgb, 255:red, 0; green, 0; blue, 0 }  ][line width=0.75]    (10.93,-3.29) .. controls (6.95,-1.4) and (3.31,-0.3) .. (0,0) .. controls (3.31,0.3) and (6.95,1.4) .. (10.93,3.29)   ;
\draw    (80.6,224.8) -- (92.22,224.6) ;
\draw    (81,90.8) -- (90.47,91.44) ;
\draw [shift={(92.47,91.54)}, rotate = 182.94] [color={rgb, 255:red, 0; green, 0; blue, 0 }  ][line width=0.75]    (10.93,-3.29) .. controls (6.95,-1.4) and (3.31,-0.3) .. (0,0) .. controls (3.31,0.3) and (6.95,1.4) .. (10.93,3.29)   ;
\draw    (81,90.8) -- (80.84,147.5) -- (80.6,224.8) ;
\draw    (181,146.4) -- (180.77,171) ;
\draw [shift={(180.75,173)}, rotate = 270.48] [color={rgb, 255:red, 0; green, 0; blue, 0 }  ][line width=0.75]    (10.93,-3.29) .. controls (6.95,-1.4) and (3.31,-0.3) .. (0,0) .. controls (3.31,0.3) and (6.95,1.4) .. (10.93,3.29)   ;
\draw  [fill={rgb, 255:red, 255; green, 255; blue, 255 }  ,fill opacity=1 ] (282,213.75) -- (348.75,213.75) .. controls (350.56,213.75) and (352.03,215.22) .. (352.03,217.03) -- (352.03,230.47) .. controls (352.03,232.28) and (350.56,233.75) .. (348.75,233.75) -- (282,233.75) .. controls (280.19,233.75) and (278.72,232.28) .. (278.72,230.47) -- (278.72,217.03) .. controls (278.72,215.22) and (280.19,213.75) .. (282,213.75) -- cycle ;
\draw    (278.6,223.8) -- (234.45,223.8) ;
\draw [shift={(235.45,223.8)}, rotate = 360] [color={rgb, 255:red, 0; green, 0; blue, 0 }  ][line width=0.75]    (10.93,-3.29) .. controls (6.95,-1.4) and (3.31,-0.3) .. (0,0) .. controls (3.31,0.3) and (6.95,1.4) .. (10.93,3.29)   ;

\draw (151.5,92) node [align=center, inner sep=0.75pt]  [font=\footnotesize] {{\fontfamily{ptm}\selectfont Perception Camera}};
\draw (174,136.4) node [align=center, inner sep=0.75pt]  [font=\footnotesize] {{\fontfamily{ptm}\selectfont YOLOv5-s Image Processing}};
\draw (179.5,183) node [align=center, inner sep=0.75pt]  [font=\footnotesize] {{\fontfamily{ptm}\selectfont Bayesian Inference Engine}};
\draw (282.5,92) node [align=center, inner sep=0.75pt]  [font=\footnotesize] {{\fontfamily{ptm}\selectfont Reward Function}};
\draw (501,69.27) node [align=center, inner sep=0.75pt]  [font=\footnotesize] {{\fontfamily{ptm}\selectfont Feasible Action Set}};
\draw (500.75,106.36) node [align=center, inner sep=0.75pt]  [font=\footnotesize] {{\fontfamily{ptm}\selectfont Hypothetical Position and Measurement}};
\draw (501,153.75) node [align=center, inner sep=0.75pt]  [font=\footnotesize] {{\fontfamily{ptm}\selectfont Bayesian Inference Engine}};
\draw (163.83,223.75) node [align=center, inner sep=0.75pt]  [font=\footnotesize] {{\fontfamily{ptm}\selectfont Environments}};
\draw (190.13,109.65) node [anchor=north west][inner sep=0.75pt]  [font=\footnotesize]  {$\text{p}_{k}$};
\draw (134.43,107.34) node [anchor=north west][inner sep=0.75pt]  [rotate=-0.61,xslant=0] [align=left] {{\fontfamily{ptm}\selectfont {\footnotesize Images}}};
\draw (270.82,107.07) node [anchor=north west][inner sep=0.75pt]  [font=\footnotesize]  {$C_{k}$};
\draw (340,76.95) node [anchor=north west][inner sep=0.75pt]  [font=\footnotesize]  {$C_{k}$};
\draw (509.18,80.76) node [anchor=north west][inner sep=0.75pt]  [font=\footnotesize]  {$\mathcal{U}$};
\draw (507.31,123.4) node [anchor=north west][inner sep=0.75pt]  [font=\footnotesize]  {$\text{p}_{k+1|k} ,\hat{\mathcal{C}}_{k+1|k}$};
\draw (506.54,164.71) node [anchor=north west][inner sep=0.75pt]  [font=\footnotesize]  {$\text{p}_{k+1|k} ,\theta_{k+1|k}$};
\draw (513.96,219.28) node [anchor=north west][inner sep=0.75pt]  [font=\footnotesize]  {$u_{k}$};
\draw (248.04,206.85) node [anchor=north west][inner sep=0.75pt]  [font=\footnotesize]  {$\text{p}_{k+1}$};
\draw (161.8,53.8) node [anchor=north west][inner sep=0.75pt]  [font=\large] [align=left] {{\fontfamily{ptm}\selectfont {\Large  }DCEE-based GOCS}};
\draw (294.3,165.9) node [anchor=north west][inner sep=0.75pt]  [font=\footnotesize]  {$\mathrm{p}_{k} ,\theta _{k}$};
\draw (191.2,155.4) node [anchor=north west][inner sep=0.75pt]  [font=\footnotesize]  {$\text{p}_{k} ,\ \mathcal{C}_{k}$};
\draw (315.37,223.75) node [align=center, inner sep=0.75pt]  [font=\footnotesize] {{\fontfamily{ptm}\selectfont System}};
\draw (505,189) node [align=center, inner sep=0.75pt]  [font=\footnotesize] {{\fontfamily{ptm}\selectfont Cost Function with Exploration-Exploitation Balance}};

\end{tikzpicture}

%% file: Figures/Demo_s1_AOD_DCEE.tex
\begin{tikzpicture}
\node at (0, 0) { \includegraphics[trim=10 20 10 20, clip, width=0.5 \textwidth]{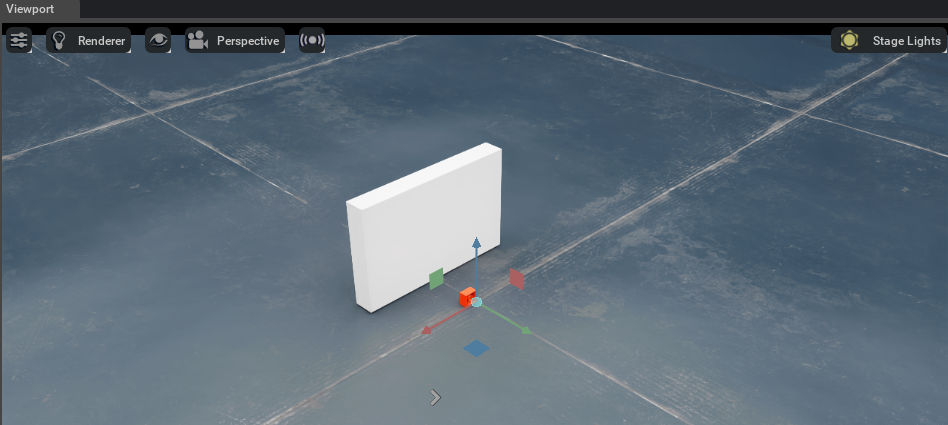} };
\node at (-2.2, 0.05) { 
\begin{tikzpicture}
 \robotArm[
config={q1=100,q2=-80},geometry={a1=01,a2=1},
end effector/.cd,
draw link/.code={
\draw[line cap=round, double=lightgray,
double distance=4mm]
(0,0) to[left] (2, 0.8);},
draw end effector/.code={
\path[link style]
(1.8, 0.8) arc (180:0:0.2) -- (2.1, 0.4)
-- (1.9, 0.4) -- cycle;
\path[link style]
(2.05,0.4) rectangle (1.95,0.0);
\fill (2,0) circle (0.05);}
]{2}
\end{tikzpicture}
    };
    \node at (-.35, 0.2){
    \begin{tikzpicture}[scale=0.3]
    \def\rotationAngle{-60} %
    
    \begin{scope}[rotate=\rotationAngle]
        
        \draw[thick] (-0.5,-0.3) rectangle (0.5,0.3); 
        \draw[thick] (0.5,0) circle (0.1); %
        
        \draw[thick, dashed] (0.5,0) -- (4,2);  %
        \draw[thick, dashed] (0.5,0) -- (4,-2); %
        
        \fill[blue!20, opacity=0.3] (0.5,0) -- (4,-2) arc [start angle=-32.57, end angle=26.57, radius=4cm] -- cycle;

    \end{scope}
\end{tikzpicture}
    };
\end{tikzpicture}

%% file: Figures/path_s1_aod_dcee.tex
\begin{tikzpicture}
    \node at (0, 0) {
        \includegraphics[trim=20 2 22 10, clip, width=0.5 \textwidth]{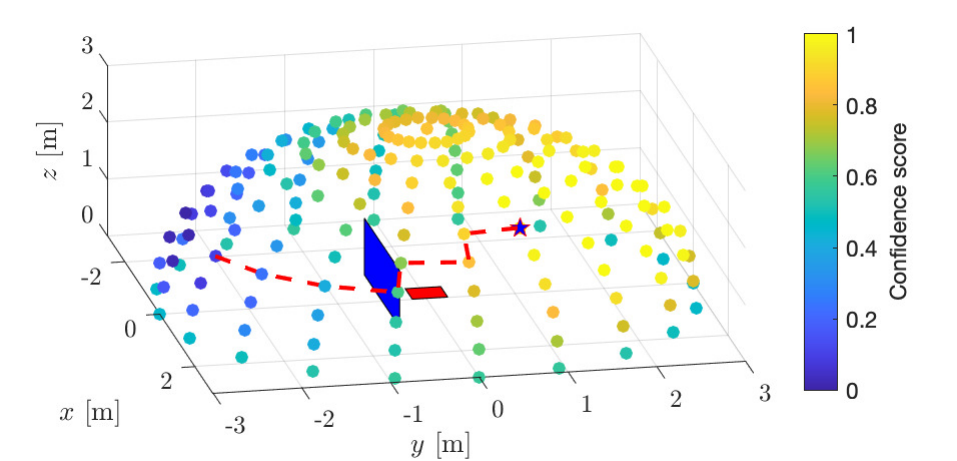}
    };
    \node at (-0.5, -0.54){
\begin{tikzpicture}
    \begin{axis}[
        view={120}{30}, %
        hide axis, %
        xmin=0, xmax=3,
        ymin=0, ymax=3,
        zmin=0, zmax=3
    ]
        \def\a{0.2}
        
        \addplot3 [
            fill=red, opacity=0.8
        ] coordinates {
            (0,0,0) (0+\a,0,0) (0+\a,0+\a,0) (0,0+\a,0) (0,0,0)
        };
        \addplot3 [
            fill=red, opacity=0.8
        ] coordinates {
            (0,0,0) (0,0+\a,0) (0,0+\a,0+\a) (0,0,0+\a) (0,0,0)
        };
        \addplot3 [
            fill=red, opacity=0.8
        ] coordinates {
            (0,0,0) (0+\a,0,0) (0+\a,0,0+\a) (0,0,0+\a) (0,0,0)
        };
    \end{axis}
\end{tikzpicture}

    };
\end{tikzpicture}

%% file: Figures/ros2_system_diagram.tex
\begin{tikzpicture}[node distance=1cm, scale=0.85]

\node (env) [block, minimum height=3.2cm, minimum width=4.5cm] {Environment};
\node (lego) [subblock] at ([yshift=0.9cm]env.center) {LEGO Brick};
\node (obstacle) [subblock] at ([yshift=-0.9cm]env.center) {Obstacle (Rectangular Box)};

\node (robot) [block, right of=env, xshift=4cm, minimum height=2cm, minimum width=4cm] {UR5e Robot Arm  (with end-mounted RealSense Camera)};

\node (ros2) [topic, right of=robot, xshift=3.5cm] {ROS2 Topics};

\node (pc) [block, right of=ros2, xshift=3.5cm, minimum height=3.2cm] {Host PC};
\node (yolo) [subblock] at ([yshift=0.9cm]pc.center) {YOLOv5 Detector};
\node (dcee) [subblock] at ([yshift=-0.9cm]pc.center) {DCEE-based GOCS};

\path [draw, thick, dashed, -latex] (env) -- (robot);   %
\path [line] (robot) -- (ros2);  %
\path [line] (ros2) -- (pc);     %

\end{tikzpicture}

%% file: Figures/expt_setting_xyz.tex
\begin{tikzpicture}[thick,->]

\node at (-0.5,1.2) {\includegraphics[width=0.5\textwidth]{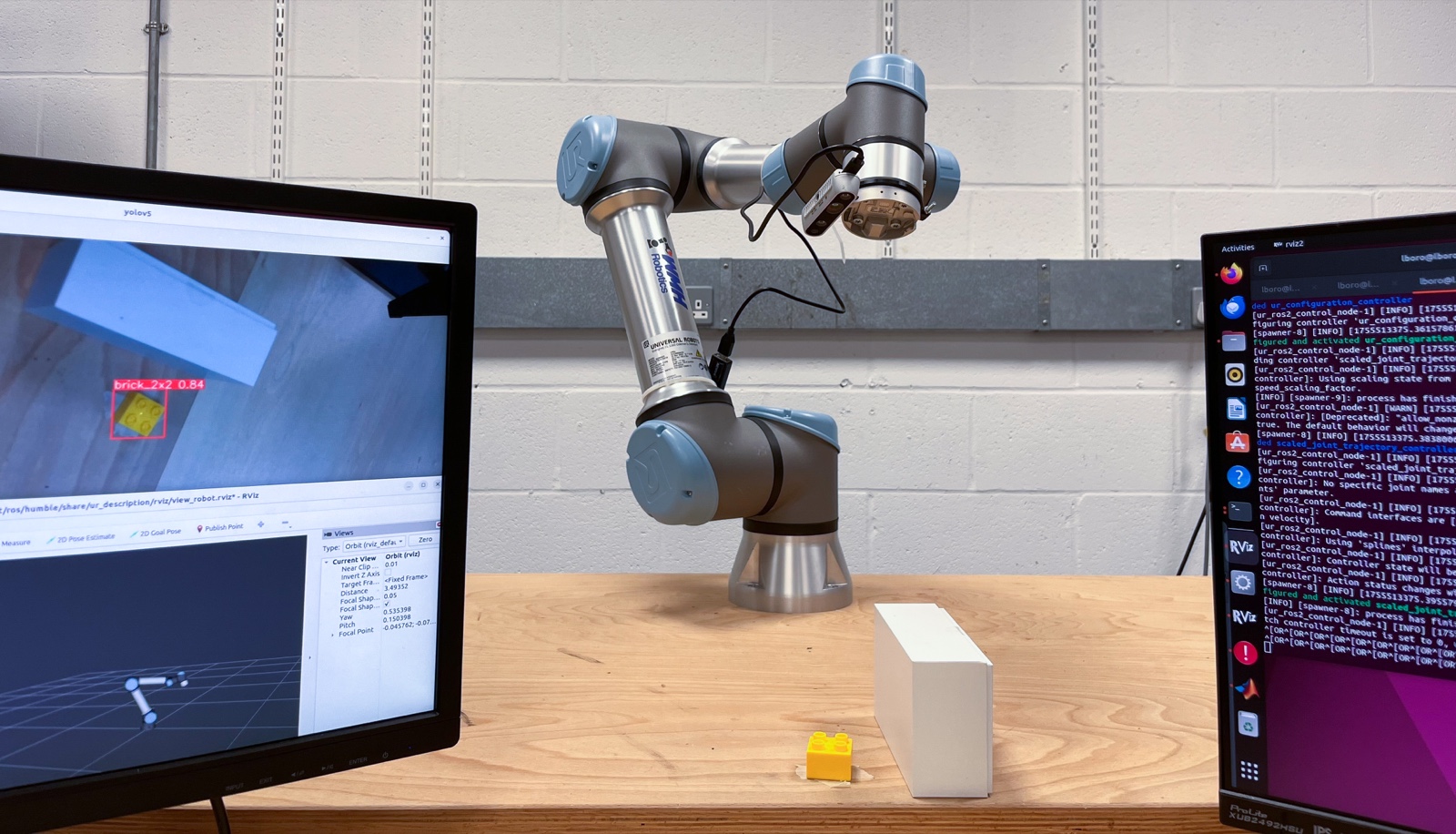}};

\draw[red] (0,0) -- (1,0) node[anchor=north]{$x$};

\draw[green!70!black] (0,0) -- (0.5,0.5) node[anchor=north]{$y$};

\draw[blue] (0,0) -- (0,1.3) node[anchor=south]{$z$};

\end{tikzpicture}

%% file: Figures/peg_s4/Demo_s4_peg_setup.tex
\begin{tikzpicture}
\node at (0, 0) { \includegraphics[width=0.5\textwidth]{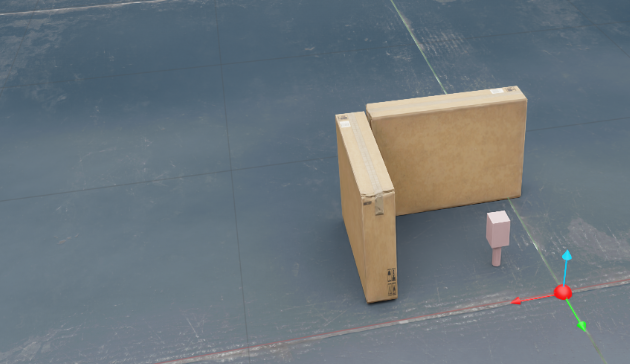} };
\clip (-4.54,-2.786) rectangle (4.534,2.786);
\node at (-0.3, 0.5) {
\begin{tikzpicture}
 \robotArm[
config={q1=100,q2=-80},geometry={a1=01,a2=1},
end effector/.cd,
draw link/.code={
\draw[line cap=round, double=lightgray,
double distance=4mm]
(0,0) to[left] (2, 0.8);},
draw end effector/.code={
\path[link style]
(1.8, 0.8) arc (180:0:0.2) -- (2.1, 0.4)
-- (1.9, 0.4) -- cycle;
\path[link style]
(2.05,0.4) rectangle (1.95,0.0);
\fill (2,0) circle (0.05);}
]{2}
\end{tikzpicture}
    };
    \node at (1.6, .50){
    \begin{tikzpicture}[scale=0.3]
    \def\rotationAngle{-60} %

    \begin{scope}[rotate=\rotationAngle]

        \draw[thick] (-0.5,-0.3) rectangle (0.5,0.3);
        \draw[thick] (0.5,0) circle (0.1); %

        \draw[thick, dashed] (0.5,0) -- (4,2);  %
        \draw[thick, dashed] (0.5,0) -- (4,-2); %

        \fill[blue!20, opacity=0.3] (0.5,0) -- (4,-2) arc [start angle=-32.57, end angle=26.57, radius=4cm] -- cycle;

    \end{scope}
\end{tikzpicture}
    };
\end{tikzpicture}